\documentclass[a4paper,fleqn,usenatbib]{mnras}

\usepackage{newtxtext,newtxmath}

\usepackage[T1]{fontenc}
\usepackage{ae,aecompl}

\usepackage{graphicx}	
\usepackage{amsmath}	
\usepackage{amssymb}	

\title[]{A search for transiting planets around FGKM dwarfs and sub-giants in the TESS Full Frame Images of the Southern ecliptic hemisphere}

\author[Montalto et al.]{
M. Montalto$^{1,2}$\thanks{E-mail: marco.montalto@unipd.it},
L. Borsato$^{2}$,
V. Granata$^{1,2}$,
G. Lacedelli$^{1,2}$,
L. Malavolta$^{1}$,
\newauthor
E. E. Manthopoulou$^{1,2}$,
D. Nardiello$^{2,3}$,
V. Nascimbeni$^{1,2}$,
G. Piotto$^{1,2}$
\\
$^{1}$Dipartimento di Fisica e Astronomia "Galileo Galilei", Universit\'a di Padova, Vicolo dell'Osservatorio 3, Padova IT-35122, Italy\\
$^{2}$Istituto Nazionale di Astrofisica - Osservatorio Astronomico di Padova, Vicolo dell'Osservatorio 5, 35122, Padova, Italy\\
$^{3}$ Aix Marseille Univ, CNRS, CNES, LAM, Marseille, France\\
}

\date{}

\pubyear{2020}

\begin{document}
\label{firstpage}
\pagerange{\pageref{firstpage}--\pageref{lastpage}}
\maketitle

\begin{abstract}
In this work, we present the analysis of 976 814 FGKM dwarf and sub-giant stars in the TESS Full Frame Images (FFIs) of the Southern ecliptic hemisphere. We present a new pipeline, DIAmante, developed to extract optimized, multi-sector photometry from TESS FFIs and a classifier, based on the Random Forest technique, trained to discriminate plausible transiting planetary candidates from common false positives. A new statistical model was developed to provide the probability of correct identification of the source of variability.  
We restricted the planet search to the stars located in the least crowded regions of the sky 
and identified 396 transiting planetary candidates among which 252 are new detections. The candidates' radius distribution ranges between 
1 R$\rm_{\oplus}$ and 2.6 R$\rm_J$ with median value of 1 R$\rm_J$ and the period distribution ranges between 0.25 days and 105 days with median value of 3.8 days. The sample contains four long period candidates (P>50 days) one of which is new and 64 candidates with periods between 10 and 50 days (42 new ones). In the small planet radius domain (R<4 R$\rm_{\oplus}$) we found 39 candidates among which 15 are new detections. Additionally, we present 15 single transit events (14 new ones), a new candidate multi-planetary system and a novel candidate around a known TOI. By using {\it Gaia} dynamical constraints we found that 70 objects show evidence of binarity. We release a catalog of the objects we analyzed and the corresponding lightcurves and diagnostic figures through the MAST and ExoFOP portals.
\end{abstract}




\section{Introduction}
\label{sec:introduction}

The {\it Transiting Exoplanet Survey Telescope} \citep[{\it TESS}, ][]{ricker2014} was launched on April 18, 2018 aboard a SpaceX Falcon 9 rocket out of Cape Canaveral.
 With its array of four cameras, covering in total 24 degrees by 96 degrees of the sky during each pointing, the satellite is delivering both short cadence (2 min) imagettes on pre-selected targets and Full Frame Images (FFIs) with a cadence of 30 min. Each pointing corresponds to a sector of the sky and it is observed for a period of about 27 days, after which the satellite moves to the next one.

\noindent
This is the first space-based transit search mission monitoring nearly the entire celestial sphere and focusing in particular on bright stars. Its eminent precursors were {\it CoRoT} \citep{baglin2003} and {\it Kepler} \citep{borycki2005} which monitored comparatively smaller, selected regions of the sky, targeting typically fainter stars than {\it TESS}, and {\it K2} \citep{howell2014} which employed the same {\it Kepler} satellite to analyze 21 ecliptic fields. Focusing on bright stars is crucial both to facilitate the confirmation of transiting candidates and to enable a galore of follow-up, characterization studies on confirmed planets. Very recently, the CHaracterising ExOPlanet Satellite \citep[\textit{CHEOPS}][]{broeg2013} became operational. Planets around bright stars are primary targets for {\it CHEOPS} to measure their radii with unprecedented precision and probe their internal structure thanks to concomitant, ground-based, high-precision spectroscopic observations which permit to derive their masses. These objects are also of great importance for atmospheric characterization studies for example through transmission or emission spectroscopy and will be in particular prime targets for {\it JWST} \citep{gardner2006} and {\it ARIEL} \citep{tinetti2018}. 

The importance of {\it TESS} goes well beyond its primary mission goals. It will be also relevant for a wealth of different astrophysical studies. In this work, we introduce our project which consists to derive accurate, space-based photometry for all the dwarf and sub-giant stars of the sky satisfying our constraints (Sec.~\ref{sec:stellar_sample}). Such an effort will be important for the characterization of the variability properties of these objects and will be relevant also for the prioritization of targets of the future planet-hunting mission {\it PLATO} \citep{rauer2014}. In particular, thanks to {\it TESS} FFIs it is possible to derive the photometry of unbiased lists of stars and extract information on: multiplicity (from the detection of eclipsing binaries down to small transiting planets), activity, pulsations, rotational periods are all  physical quantities that can be obtained from {\it TESS} lightcurves. This can be done for millions of stars spread across the entire sky and with unprecedented photometric precision. In this work, we explore the potential of {\it TESS} FFIs to detect transiting planets, analyzing a carefully selected sample of FGKM  dwarf and sub-giant stars in the Southern ecliptic hemisphere. 

\noindent
At the time of writing, {\it TESS} has nearly completed its core mission, monitoring both the Southern and the Northern ecliptic hemispheres and it is moving toward its already announced extended mission, where both the number of short cadence targets and FFIs will be increased. So far, 51 confirmed planets and 2040 transiting candidates have been announced\footnote{A useful list of \textit{TESS} publications can be found at \url{https://heasarc.gsfc.nasa.gov/docs/tess/tpub.html}} \citep[e.g.][]{huang2018,vanderspek2019,wang2019,gandolfi2018,trifonov2019,king2019,shporer2019,bouma2019a}. 

\noindent
In Sec.\ref{sec:observations}, we describe the observations and in Sec.\ref{sec:stellar_sample} the stellar sample we analyzed. We present the new pipeline we used to extract the photometry from {\it TESS} FFIs in Sec.~\ref{sec:differential_image_photometry}.
In Sec.~\ref{sec:masking}-Sec.~\ref{sec:normalization} we discuss our post-reduction analysis and in Sec.~\ref{sec:photometric_precision} the photometric precision we achieved. In Sec.~\ref{sec:transit_detection}, we present the transit search approach and in Sec.~\ref{sec:classification} we discuss the classification algorithm we adopted to discriminate plausible transiting planets from several common false positives. The centroid motion algorithm is presented Sec.~\ref{sec:centroid_motion} and the selection of transiting candidates in Sec.\ref{sec:selection_planetary_candidates}. The results of the analysis are illustrated in Sec.~\ref{sec:results} and the discussion of the results in Sec.~\ref{sec:discussion}. We finally draw our conclusions in Sec.~\ref{sec:conclusions}.

\begin{figure*}
\includegraphics[width=0.65\columnwidth]{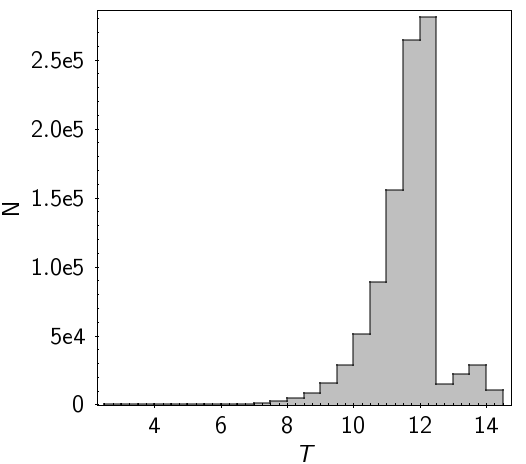}
\includegraphics[width=0.65\columnwidth]{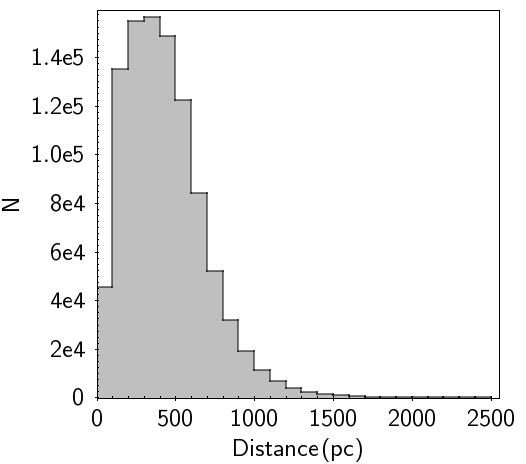}
\includegraphics[width=0.65\columnwidth]{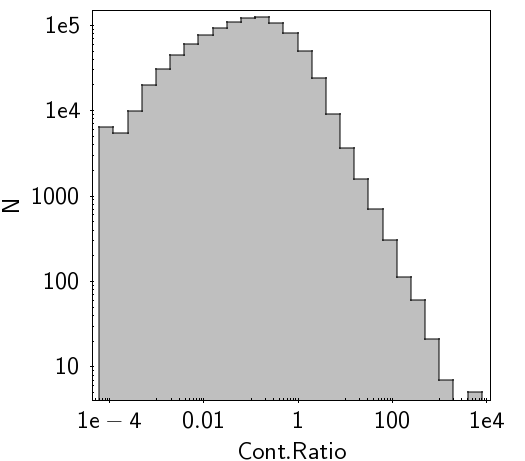}
\includegraphics[width=0.65\columnwidth]{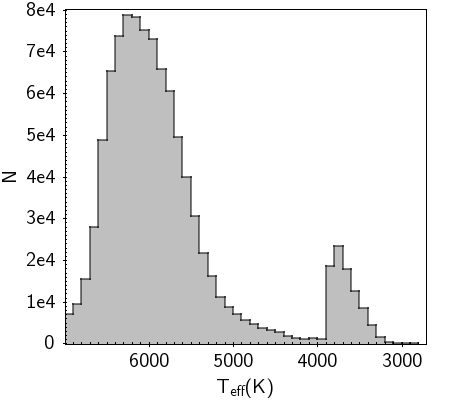}
\includegraphics[width=0.65\columnwidth]{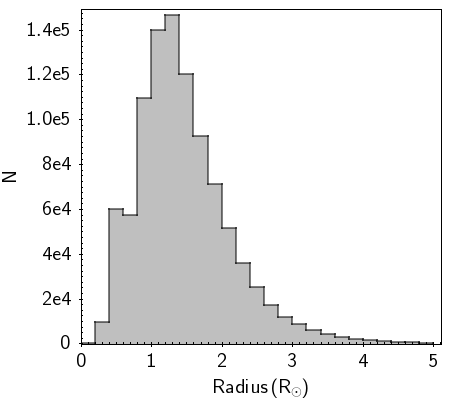}
\includegraphics[width=0.65\columnwidth]{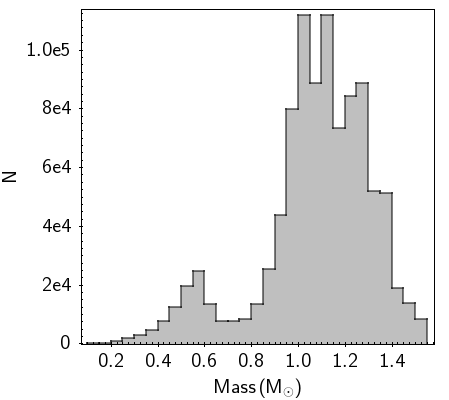}
\caption{\textit{Top, from the left:} 
distribution of \textit{TESS} magnitudes, distances and contamination ratios for the stars analyzed in this work.
\textit{Bottom, from the left:} 
distribution of stellar effective temperatures, radii and masses. Values of stellar parameters
are taken from \citet{stassun2019}.}
\label{fig:sample}
\end{figure*}

\section{Observations}
\label{sec:observations}
 
 In this work, we analyzed the {\it TESS} FFIs delivered by the satellite during its first year of operation. The images cover the Southern ecliptic hemisphere. Each sector is imaged by four {\it TESS} cameras which are composed by four {\it TESS} CCDs. Therefore, each cadence image delivered by {\it TESS} is made up of sixteen FFIs. The dataset we analyzed contains 15347 epochs which correspond to 245 552 FFIs images. The first observation started on July 25, 2018 at UT=19:29:42.708 
 and the last observation on July 17, 2019 at UT=19:59:29.974. We started our analysis from the calibrated FFIs. \textit{TESS} cameras are read out every 2 seconds and the resulting images are stored into 30 minutes exposures. Cosmic rays are mitigated by an on-board algorithm \citep{vanderspek2019} and sent to the ground where they are processed by the Science Processing Operations Center \citep[SPOC,][]{jenkings2016}.
 The SPOC performs traditional  CCD  data  reduction steps (e.g. correction for bias, dark current and flat field), as well as {\it TESS}-specific corrections (removing smear signals resulting from the lack of a shutter on the cameras). The  resulting  science  data  products are  described  by \cite{tenenbaum2018}.

\begin{figure*}
\includegraphics[width=\columnwidth]{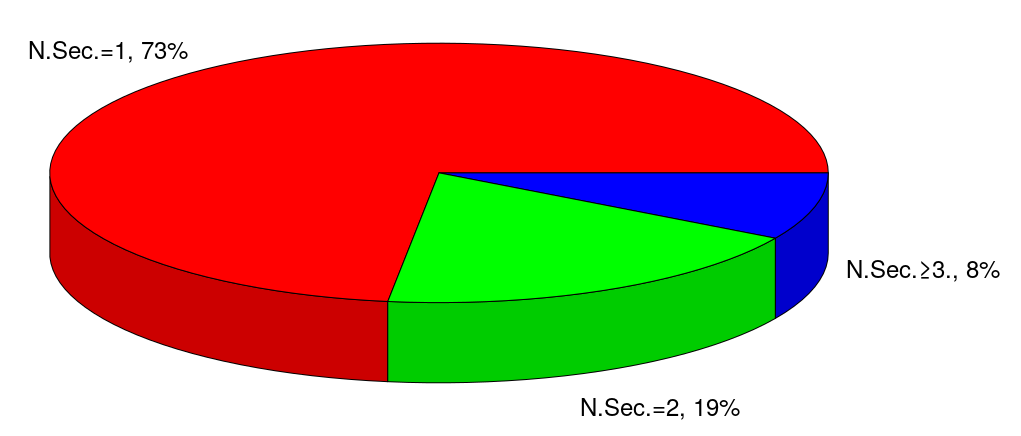}	
\includegraphics[width=\columnwidth]{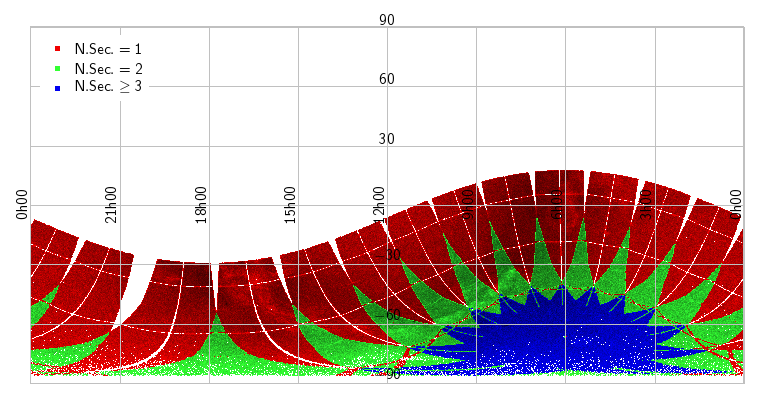}
\caption{
\textit{Left:} pie chart representing the percentual number of targets observed in one, two or more than two sectors.
\textit{Right:} Spatial distribution of our selected targets across the celestial sphere in an equatorial
coordinate reference system. Colors denote the number of sectors where a star was measured, as specified in the legend. 
}
\label{fig:FOV}
\end{figure*}

\begin{table}
	\centering
	\caption{Number of stars (N. stars), percentual number of stars and number of sectors (N. sectors) where the stars are imaged.}
	\label{tab:sectors}
	\begin{tabular}{rrr} 
		\hline
		N. stars & $\%$ & N. sectors \\
		\hline        
		711981 & 72.9 & 1 \\
		182383 & 18.7 & 2 \\
		34368  & 3.5 & 3 \\
		 6464 & 0.7 & 4 \\
	     3379 & 0.3 & 5 \\
	     4084 & 0.4 & 6 \\
	     2835 & 0.3 & 7 \\
	      1503 & 0.2 & 8 \\
	      1796 & 0.2 & 9 \\
	      2082 & 0.2 & 10 \\
	      4580 & 0.5 & 11 \\
	     10338  & 1.0 & 12 \\
	     11021 &  1.1 & 13 \\
	     \hline
    \end{tabular}        
\end{table}

\section{Stellar sample}
\label{sec:stellar_sample}

The stellar sample we analyzed was built from the \textit{Gaia} DR2 all-sky catalog. It is a sample of FGKM (limited to F5) dwarf and sub-giant stars. The selection was done in the absolute, intrinsic color magnitude diagram by imposing conditions on the absolute magnitude and colors. Stellar distances were derived from \cite{bailer2018} and reddening from the interpolation of \cite{lallement2018} 3D-maps. For FGK dwarfs and sub-giants the selection was limited to $V<13$. For dwarfs we imposed: 0.42$\,<\,$\textit{(B-V)}${_0}<$1.38, M$_{V,0}\ge\,$5\,\textit{(B-V)}$_{0}$+0.4 and M$_{V,0}<$5\textit{(B-V)}$_{0}$+3.5 while for sub-giants: M$_{V,0}<$5\textit{(B-V)}$_0$+0.4 and M$_{V,0}>$5\textit{(B-V)}$_0$-2 in the color range between 0.42$<$\textit{(B-V)}$_0\leq$0.8 and M$_{V,0}<$4.5 and M$_{V,0}>$5\textit{(B-V)}$_0$-2 in between 0.8$<$\textit{(B-V)}$_0\leq$1.0.
Here M$_{V,0}$ is the absolute, intrinsic magnitude in the $V$-band and \textit{(B-V)}$_0$ is the absolute \textit{(B-V)} color. Optical \textit{B} and \textit{V} band photometry was derived from \textit{Gaia} colors using calibration relations in \cite{evans2018b}. For M-dwarfs, the selection was performed down to $V\,\leq\,$16 and distance$\,<\,$600 pc, imposing ($G\rm_{BP}$-$G\rm_{RP}$)$_0\geq1.84$ and M$\rm_{G,0}>$2.334($G\rm_{BP}$-$G\rm_{RP}$)$_0$+2.259, where M$\rm_{G,0}$ is the absolute, intrinsic magnitude in the $G$-band and ($G\rm_{BP}$-$G\rm_{RP})_0$ is the absolute intrinsic ($G\rm_{BP}$-$G\rm_{RP})$ color. The optical \textit{V}-band photometry was derived in this case as a function of the \textit{Gaia} color using a custom calibrated relation. Further details on the catalog construction will be provided in Montalto et. al. (in preparation). This catalog was restricted to stars falling in the footprint of \textit{TESS} CCDs of the first thirteen sectors and matched with the \textit{TESS} Input Catalog \citep[v8, ][]{stassun2019} restricted to stars with T$\rm_{eff}<$7000 K, log g$>3$ and TIC $V<13$ for FGK dwarfs and sub-giants, and to stars with T$\rm_{eff}<3870$ K, log g$>3$ and TIC $V<16$ for M-dwarfs. From the TIC catalog, we extracted the stellar parameters used in this work.
The number of stars and the number of sectors in which they are observed are reported in Table~\ref{tab:sectors}. In Fig.~\ref{fig:sample}, we show the distributions of
\textit{TESS} magnitudes, stellar distances, contamination
ratios, effective temperatures, stellar radii and masses of the selected stars.
Figure~\ref{fig:FOV} (left panel)
shows a pie chart representing the percentual number of targets which are observed in one, two and more than two sectors and the targets' distribution across the celestial sphere (right panel). The sample contains 889 411 FGK dwarfs and 87 403 M-dwarfs for a grand total of 976 814 stars.

\section{Data reduction}
\label{sec:differential_image_photometry}

Several approaches have been presented so far for the analysis of {\it TESS} FFIs \citep[e.g.][]{oelkers2018,bouma2019,feinstein2019,handberg2019,nardiello2019,nardiello2020}. Here we employ a method based on the
difference image analysis  \citep[DIA][]{alard1998,alard2000,bramich2008,miller2008}.
Difference imaging permits a very efficient subtraction of all constant sources in the field and therefore
reduces the impact of contaminants on the target's photometry and
permits a more accurate estimate of the sky background with respect to
simple aperture photometry. Moreover, the subtracted images can also be 
exploited during the centroid motion analysis, as explained in Sect.~\ref{sec:centroid_motion}.
The technique is based on the subtraction of a high S/N reference image (R) of a stellar field to a target image (I) after convolving the reference with an optimal Kernel (K) to match the intensity, the background (B) and the PSF of the target image. As a kernel we chose the delta function basis kernel \citep{miller2008} and experimented with different kernel  dimensions between 3 to 5 pix and constant, first and second order expansions. We found that in most cases a constant kernel of dimension 3 $\times$ 3 pix provided the best compromise between accuracy of the subtraction and efficiency. Moreover, as demonstrated by \cite{miller2008}, such a kernel has the ability to compensate for small drifts of the target and reference image directly during the kernel solution step. This is an attractive feature because it permits to avoid the usual step of image registration and ensures a perfect flux conservation. In our approach we therefore convolve the reference image with the optimal kernel and extract the photometry on the reference system of the target image by using the WCS solution embedded in each FFI. Moreover, we also solve for the differential background between the reference and the target image simultaneously with the kernel, as explained below. 
The best kernel solution is then found by solving the equation

\begin{equation}
\rm R\otimes\,K+B=I    
\end{equation}

\noindent
where the symbol $\otimes$ denotes the convolution operation and B is the differential background model which we initially assumed constant. 

The kernel K was expanded in a set of M$^2$ delta basis kernels K$\rm_{p,q}$ which are M$\times$M matrixes whose elements are

\[
  \rm K_{p,q}(i,j)=\begin{cases}
              \rm 1\,\,if\,\,(i=p\,\land\,j=q)\\
              \rm 0\,\,if\,\,(i\neq p\,\lor\,j\neq q)
          \end{cases}
\]

\noindent
then Eq.~1 is rewritten as

\begin{equation}
\rm \sum_{p=1}^{M}\sum_{q=1}^{M}\,A_{p,q}\,(R\otimes\,K_{p,q})+B=I   
\end{equation}

\noindent
where A$\rm_{p,q}$ are the Kernel coefficients.  One of the peculiar characteristics of {\it TESS} FFIs is that they present rather erratic background variations depending on the boresight angle between each camera and the Sun and the Moon directions. Because of this, we then constructed a more accurate differential background model. We considered the first iteration subtracted image and filtered it to remove flux variations' high spacial frequencies which are typically associated with the residuals of stellar sources. To do this, we first determined the flux's dispersion ($\sigma$) of the entire subtracted image. We then calculated, for each pixel, the absolute flux differences ($\rm |df|$) between that pixel and all surrounding pixels. If the condition $\rm |df|>\sigma$ was met for at least one of these flux differences, the pixel was masked. We then set each image pixel's value equal to the average flux calculated in a square region centered on the pixel (considering only the unfiltered pixels in the averaging process). After some tests, we adopted a value of 20 pix for the dimension of the box smoothing region. Such radius allows to model both small and large scale {\it TESS} background variability. The filtered and smoothed differential background model (B$\rm_{im}$) was then simultaneously fit with the Kernel by replacing the constant term B in Eq.~1 with a first order expansion

\begin{equation*}
\rm B^{\prime}=B_1\times B_{im}\,+\,B_2.
\end{equation*}

\begin{figure*}
\includegraphics[width=16.5cm]{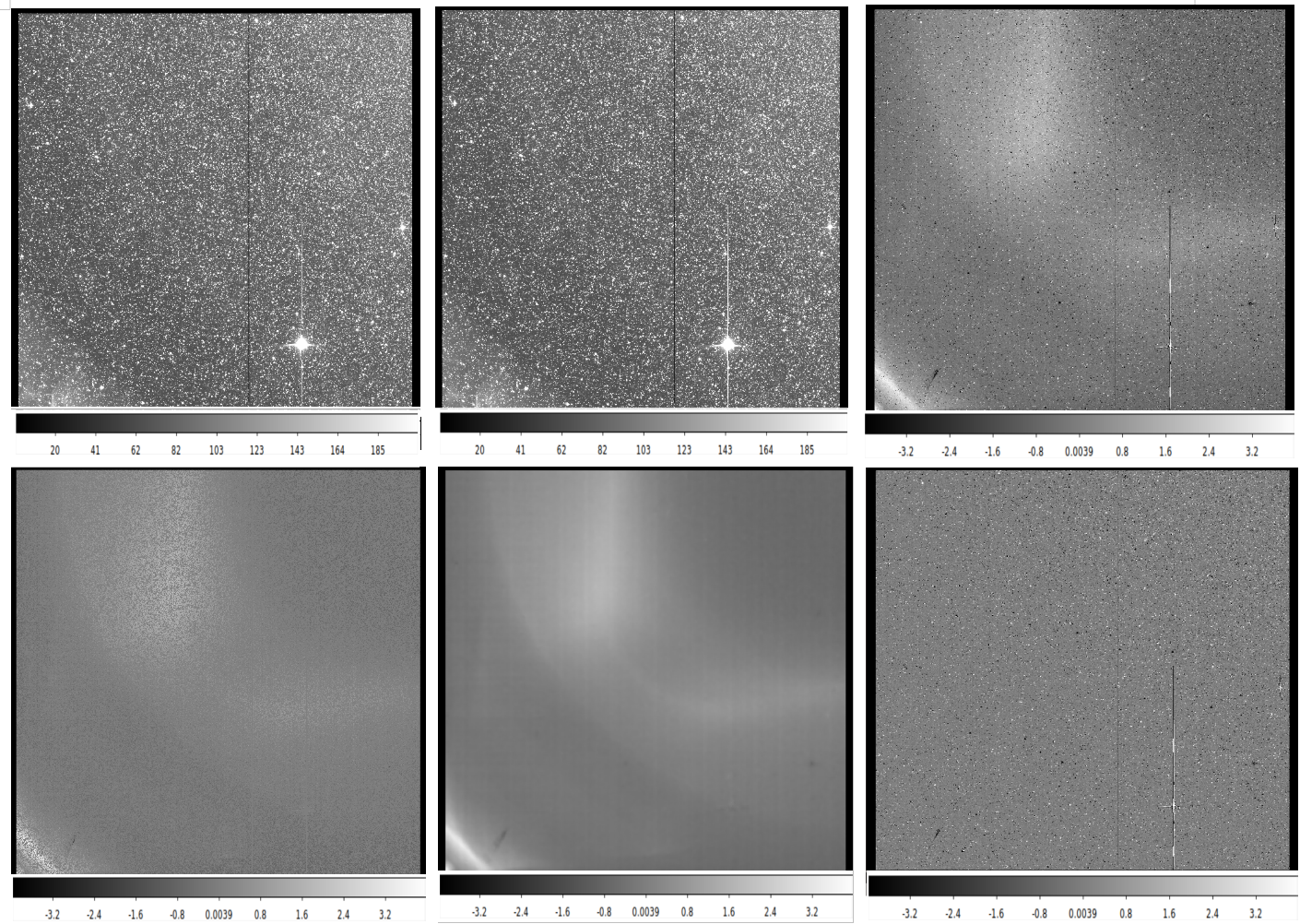}
\caption{Overview of the reduction process. In the the top raw from the left side: original image, reference image, first iteration subtracted image. On the bottom raw, from the left side: filtered subtracted image, filtered and smoothed differential background model and final subtracted image. The image represented is tess2019022075936-s0007-3-3-0131-s\_ffic.fits of Sector 7, camera 3, CCD 3.
}
\label{fig:image_reduction}
\end{figure*}

\noindent
Once the optimal Kernel solution was found, the convolved reference image and the background terms were subtracted from the target image to form the final subtracted image. Fig.~\ref{fig:image_reduction} shows an overview of the reduction process for the image tess2019022075936-s0007-3-3-0131-s\_ffic.fits of Sector 7, camera 3, CCD 3. 
In the video at \url{https://youtu.be/EpuIPCcgTo8}
we also show the full sequence of 1087 subtracted images for Sector 7, Camera 3, CCD 3\footnote{We excluded six images affected by momentum dumps.}. The frame rate is equal to 12 images/sec. At the beginning of the video (0:00-0:01) and at (0:46-0:47) vertical "straps"  are visible.
 Towards the end of the sequence (1:09) a moving object entered the camera FOV from the top right corner.

\subsection{Reference image}
\label{sec:reference}

From the WCS solutions of each image we deduced the zero order shift of each image with respect to a reference astrometric image. This image was usually chosen as the first frame for which a WCS solution  was reported and that did not have any obvious defect. For each image, we then considered the relative offset in the X and Y coordinates between that image and all other images in the set and determined the minimum relative offset. 
If such minimum offset was larger than 0.003 pix we excluded the image from the set of images used to build the reference.
This procedure served to eliminate images with relatively big isolated offsets with respect to the others. From the remaining set we then produced a median stack image, which we used as reference frame for our analysis. We also calculated the median offset between all the selected frames used to build the reference and the astrometric reference frame, which we then used to project our masterlist on the reference frame to extract the reference frame's flux.

\subsection{Photometry}
\label{sec:photometry}

\noindent
We extracted the photometry on a set of two concentric circular apertures with radii of 1 pix and 2 pix. The aperture was centered on the {\it Gaia} catalog predicted positions of our targets, corrected by proper motions at the beginning of each {\it TESS} sector campaign whenever the relative proper motion error on each coordinate was smaller than 10$\%$. The conversion between the sky coordinates and the image coordinates was performed using the WCS solution embedded in each image using the task \texttt{sky2xy} of the \texttt{WCSTools} package \citep{mink1997}.
We checked the quality of the astrometric solution for a sample of representative images.
We considered the centroid positions derived from the WCS solution as initial guesses for a PSF fitting algorithm, by means of which we further refined the centroid coordinates. We used in particular the empirical PSFs provided by the {\it TESS} team\footnote{https://archive.stsci.edu/missions/tess/models/prf\_fitsfiles/}. We then compared the PSF centroid coordinates with the WCS solution coordinates and obtained the following median differences: |{\bf dr}|=0.03 pix with standard deviation $\rm \sigma_{|{\bf dr}|}=$0.18 pix . Given that our aperture radii are either 1 pix or 2 pix we considered this difference acceptable and adopted the WCS solution coordinates.

Denoting with f$\rm_{i,j}$(S) the flux of the subtracted image in the (i,j) image pixel and with f(I)  and f(R) the integrated flux of an object in the target and reference image we have

\begin{equation}
\label{eq:flux}
\rm f(I)=C_s\left(f(R)+\frac{1}{||K||}\,\sum_{(i,j)\in\Gamma}w(i,j)f_{i,j}(S)\right)
\end{equation}

\noindent
where the summatory was extended to all pixels included or intersecting a  given aperture $\Gamma$. The w(i,j) function is a weighting function which gives the fractional area of pixel (i,j) included in the aperture and it was calculated using the Green's theorem. The term $\rm ||K||$ is the Kernel norm which represents the flux scaling factor between the reference and the target image. The flux $\rm f(R)$ on the reference image was calculated from the {\it TESS} magnitude ({\it T}) reported in the TIC. For each CCD we calculated the median magnitude difference ($\rm \Delta\,m$) between the instrumental magnitudes of a sample of 250 calibration stars  (measured on the reference image adopting the same aperture used for the target image), and the corresponding {\it TESS} magnitudes.
These stars were selected among the most bright, non-saturated and isolated stars in each CCD.
Then the flux $\rm f(R)$ was obtained from

\begin{equation}
\rm f(R)=10^{-0.4({\it T}+\Delta m)}
\end{equation}

\noindent
The constant $\rm C_s$ in Eq.~\ref{eq:flux} accounts for sector, camera, CCD systematic zero points. Denoting with $\rm R_s=10^{-0.4(\Delta m)}$ the median 
flux ratio between the comparison stars' instrumental fluxes of the s-th sector, camera, CCD and the {\it TESS} catalog fluxes,
and with $\rm \widehat{R}$ the median of all sectors, camera, CCDs median flux ratios, the constant $\rm C_s$ can be expressed as

\begin{equation}
\rm C_s=\frac{\widehat{R}}{R_s}.
\end{equation}

\noindent
The flux f(i) in Eq.~\ref{eq:flux} is background subtracted and the background was evaluated from an annular region surrounding each target with inner radius equal to 5 pix and outer radius equal to 15 pix. The lightcurves obtained with the procedure described in this section are referred as LC0 lightcurves.

\section{Masking}
\label{sec:masking}

Bad measurements were masked to avoid the most prominent systematics. We flagged all images which had {\it TESS} quality flag different from zero. In addition we checked the kernel norm to identify possible problems with the subtracted images. We typically flagged images having kernel norm values that differ by more than a few percent the median value of a given sector, camera, CCD dataset. Such pipeline dependent flags are merged with the {\it TESS} pipeline flags and incorporated in the bitmask we released with the lightcurves.

\section{Cotrending}
\label{sec:cotrending}

Lightcurves were then cotrended to correct for systematic effects. 
The log RMS vs {\it T} magnitude diagram was interpolated with a third order polynomial, downweighting positive residuals to allow the fit to converge towards the lower bound limit of the distribution. Stars within 1.5$\sigma$ from the interpolated curve were then selected. We calculated the Pearson correlation coefficient of each lightcurve with respect to all other lightcurves in the sample, and then the median correlation coefficient. The stars were then sorted out in increasing order of the median correlation coefficient and the 50$\%$ most correlated lightcurves were selected. We then
sorted out the lightcurves in increasing order of magnitude and the first 1200 stars were used to perform a Principal Component Analysis (PCA). The resulting eigenvectors were sorted in decreasing order of their eigenvalues (eigval). We chose then the first N eigenvectors in the sorted list satisfying the condition $\rm\frac{\sum_1^N eigval}{\sum_1^{1200} eigval}>0.9$, and selected a maximum of 10 eigenvectors. Each lightcurve was then linearly decorrelated against this set of eigenvectors. The criterium described above was adopted after testing it on several lightcurves and it was found appropriate to remove the most important systematics without affecting transit detection. It is also capable to preserve short term, intra-sector variability, but it is less appropriate to preserve long-term variability. In multi-sector observations each portion of the lightcurve belonging to a given sector, camera, CCD is decorrelated against the corresponding set of eigenvectors appropriate for the same sector, camera, CCD, but completely independently from measurements acquired on other sectors. Therefore, it is difficult to preserve variability extending on times scales longer than one sector ($\sim$27 day). In any case this fact is largely unimportant in the context of this work, which is focused on the detection of transits occurring on much shorter time-scales.
The lightcurves obtained after cotrending are referred as LC1 lightcurves.

\section{Lightcurve normalization}
\label{sec:normalization}

Before searching for transits we applied a high-pass filter to enhance transit detectability. First we proceeded by averaging each lightcurve over 8 hr time intervals and then B-spline interpolated the resulting averages. Any gap in the data for which it was not possible to calculate the time interval average was skipped in the splining process. 
Then, we identified possible outliers in the splined lightcurve searching for measurements satisying the condition  
$\rm f_i<$Q1-1.5$\times$IQR or $\rm f_i>$Q3+1.5$\times$IQR, where $\rm f_i$ is the i-th flux measurement and Q1, Q3 and IQR are the first quartile, the third quartile and the inter-quartile range of the flux measurements' distribution. We then repeated the splining procedure starting from the original lightcurve but excluding the identified outliers.
This procedure was repeated ten times by shifting the averaging intervals of 0.8 hr each time. The resulting interpolating splines were then evaluated at the instant of each observation, and these final estimates were averaged together. The above procedure was repeated also after the transit search by excluding from the splining procedure also those measurements falling within the transit window. The lightcurves obtained after normalization are referred as LC2 lightcurves.

\section{Photometric precision}
\label{sec:photometric_precision}

\begin{figure*}
\includegraphics[width=18cm]{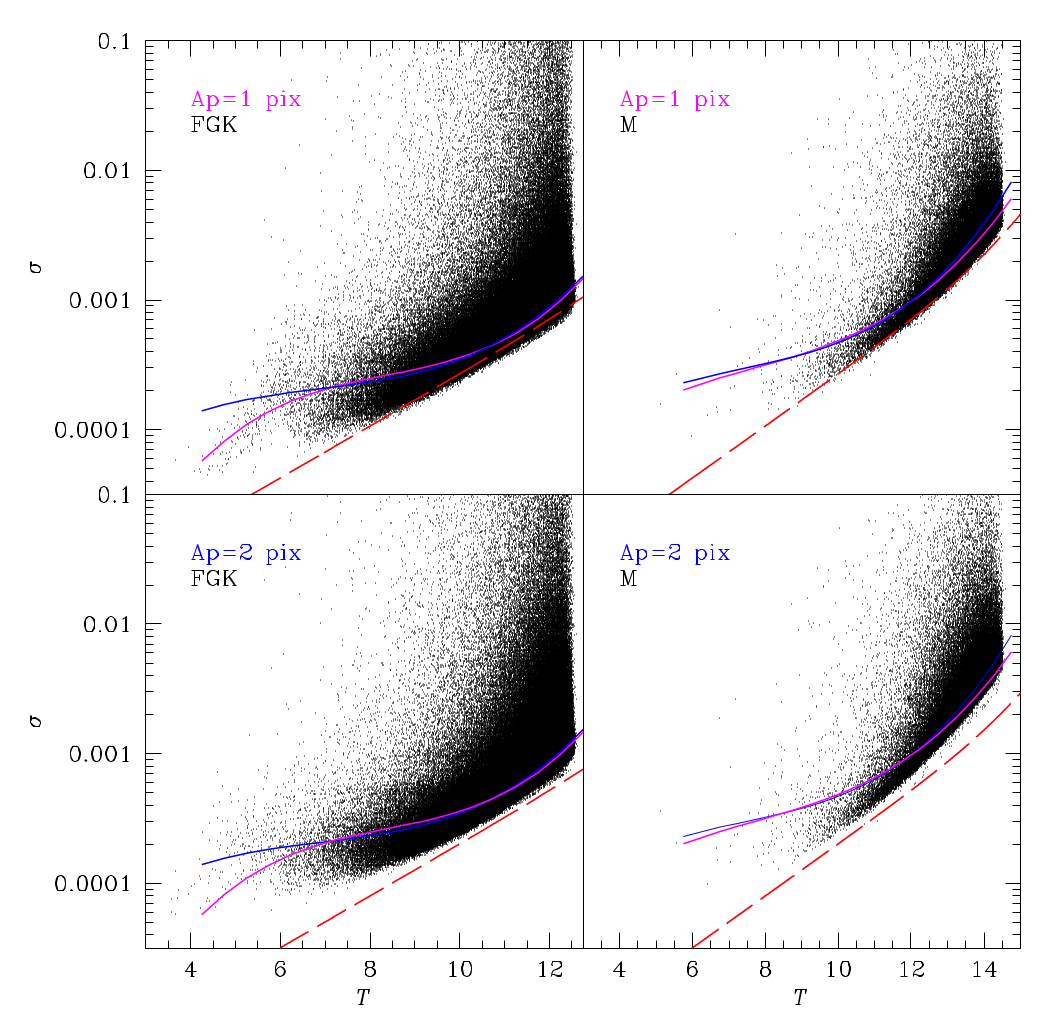}
\caption{
Photometric precision of the lightcurves for different samples and photometric apertures. The red dashed lines correspond to theoretical models, while the continuous colored lines correspond to the best fit interpolation models of observed precisions. The magenta color is related to aperture equal to 1 pix, while the blue color is related to aperture equal to 2 pix.
}
\label{fig:photometric_precision}
\end{figure*}

In Fig.~\ref{fig:photometric_precision}, we show the photometric precision ($\sigma$) achieved in the final normalized lightcurves. The photometric precision is calculated as half the difference between the 84$\rm^{th}$ and the 16$\rm^{th}$ percentiles of the cumulative distribution of the flux measurements. The diagrams in Fig.~\ref{fig:photometric_precision} display the photometric precision for the two separate samples we analyzed in this work, that is the FGK sample (left panels) and the M-dwarfs sample (right panels). They also represent the precision for the two different apertures we used and in particular for aperture$=$1 pix (upper panels, hereafter aperture 1) and for aperture$=$2 pix (lower panels, hereafter aperture 2). The observed precisions were fit with a third order polynomial function and the results are represented by the continuous magenta (aperture 1) and blue (aperture 2) lines. On the diagrams relative to a given aperture (and for the same stellar sample) we represent the best fit models of both apertures, to facilitate the comparison. The dashed-red lines are theoretical expectations and represent the photometric precision achievable for a given aperture in half-an-hour integration time, accounting for photon noise, assuming a sky equal to 150 $\mathrm{e^-\,s^{-1}}$ and a RON=8.5 $\mathrm{e^-}$.
The two apertures have in general comparable performances, but aperture 2 photometry appears to have a more stable behaviour, in particular at the bright end of the FGK magnitude sample. Aperture 1 photometry has slightly better performances for {\it TESS} magnitudes {\it T}$>$11. Considering these results we decided to perform the search for transiting planets in the full sample for all magnitudes for aperture 1 and up to {\it T}$<$11 for aperture 2.

\section{Transit detection}
\label{sec:transit_detection}

To search for planetary transits we used the box-fitting algorithm (BLS) of \citet{kovacs2002}. We searched for signals with periods in between $P_{\textrm{min}}$=0.25 days and $\Delta$T, where $\Delta$T was the total time spanned from the first to the last measurement of each lightcurve. 
The period step $\Delta$P was determined by imposing
$\Delta$P$=\frac{P}{\Delta\textrm{ T}}\epsilon$ \citep{ofir2014}. We set $\epsilon=5$min to ensure a precision equal to 5 min in the folding process between the first and the last transit in the observable window, for any tested period $P$. The fractional transit length $q$ was adjusted for any trial
period $P$ and target star radius $R_{\star}$ and mass
$M_{\star}$. In particular, we estimated the maximum transit duration $\tau_{\textrm{max}}$ at period $P$ considering a circular orbit and a 90$^{\circ}$ transiting planetary orbital inclination which gives
$\tau_{\textrm{max}}\simeq\frac{P}{\pi}\arcsin{\frac{R_{\star}}{a}}$, where $a$ is the orbital semi-major axis. We then considered fractional transit lengths $q$ in between $q_{\textrm{min}}=$0.1$\frac{\tau_{\textrm{max}}}{P}$ and $q_{\textrm{max}}=$1.1$\frac{\tau_{\textrm{max}}}{P}$. The number of bins ($n_{\textrm{bins}}$) in which each folded lightcurve was subdivided to evaluate the BLS metric was varied as well for any tested period and set equal to $n_{\textrm{bins}}=\frac{2}{q_{\textrm{min}}}$.

\section{Classification}
\label{sec:classification}

After the search for transits we applied a classification algorithm in order to identify the most promising candidates. It is known that the BLS can be sensitive to different kind of variables and/or events that mimic the shape of transiting planet signals. Some of these false positives can be usually recognized by a morphological inspection of the lightcurves. For example, it is customary to identify eclipsing binaries by analyzing odd/even transits, checking for variations of transit depths between them. Also the presence of secondary eclipses is generally an indicator of false positive signals. Ultra short period variables may be difficult to correctly cotrend/detrend and the residual signals may as well trigger the BLS. In general, the detection of these classes of objects interferes with the identification of plausible transiting planets, decreasing the efficiency of the search. It is desirable therefore to build a filter to decrease the rate of false positives. This is especially important when dealing with massive searches for planets. To tackle this problem we used a Random Forest Classifier \citep[RFC,][]{ breiman2001}. The Random Forest is an ensemble method of machine learning which uses several individual decision trees to assign a class to a given input. In order to apply it, it is necessary to define a set of \textit{features}, which can be considered as quantities related to some qualifying aspects of the signal we want to detect and of the false positive signals we want to reject. Each tree is trained on  a random subset of the input training sample, and a random subset of the input features is used at each branch of the trees to split them into sub-branches. Splits are done maximizing a given metric which optimizes correct  classification.  At the end of this process each tree provides the likelihood that a given input belongs to a certain class. Bootstrap aggregation \cite[known as {\it bagging}, e.g.][]{breiman2001} is used then to summarize the results of the single trees into a unique probabilistic estimate. The RFC is a popular classification algorithm, known for being very robust and easy to implement. It also permit a straightforward analysis of the classification results by judging for example the importance that each variable has on the classification process. RFCs have been already described in detail and applied and tested in the context of exoplanet searches in recent literature works  \citep[e.g.][]{caceres2019a,caceres2019b,schanche2019a}.

In our implementation of the RFC we focused the attention on the morphological aspects of the problem related to exoplanet transit identification. By morphological aspects we intend the fact that usually many of the features used by humans to isolate planetary transit candidates are purely morphological, in the sense that they relate one property of the lightcurve to other properties extracted from the same lightcurve. The presence of secondary eclipses, the lightcurve modulations, the morphological appearance of the primary eclipse, the odd/even transit depths, the significance of periodic signals are all qualifying morphological aspects of folded lightcurves (or of periodograms). Therefore, the set of features we defined are nearly all unitless and involve S/N estimates or similar normalized quantities evaluated at particular critical points of the lightcurves (or periodograms). Facing the problem of planetary transit classification in a purely morphological sense, it is a useful way to simplify the procedure to identify good candidates. There is evidently other important information that should be considered for a correct classification (such as dynamical information, centroid motion information for detectors like {\it TESS}), but at the first stage in the analysis,  when essentially only the lightcurves are scrutinized, a morphological analysis can already eliminate a large fraction of contaminant signals. Nevertheless, several false positives are expected to pass this test. For example, it would be evidently too pretentious to require a morphological classifier to discriminate between planetary signals and low mass eclipsing binary signals since it is known that these objects produce essentially indistinguishable transit photometry. In fact these objects can be identified only by subsequent follow-up radial velocity (or timing) analysis. 
For this reason, when evaluating the performances of the classifier, these kind of false positives should not be considered.
Moreover, feature design should also take into account the need to define quantities that can be easily and robustly calculated for any lightcurve. Feature definition is described below in Sect.~\ref{sec:feature_definition}.  

One of the problems related to the construction of a classifier of this kind is which sample of planets to consider and which classes of false positives to include in the analysis. By considering that the physics of transiting planets and of eclipsing binaries is relatively well understood (at least for what concerns the modelization of their lightcurves), in the following we resort to use  simulations to produce traning sets to feed the classifier. This has the undoubtful vantage to overcome issues related to class imbalance, where planetary signals are usually strongly underrepresented in samples drawn from the real world. However, in order to reproduce as closely as possible the conditions to which the classifier will be applied, we  selected a random, representative sample of stars from the total sample we analised in this work, and injected the artificial signals of planets or false positives into their lightcurves, as described in Sect.~\ref{sec:simulations}.

\subsection{Features' definition}
\label{sec:feature_definition}

\subsubsection{Effective S/N of the primary eclipse}

One of the most important parameters to define in the contex of transit searches is the effective S/N of the primary eclipse, as already recognized by \cite{kovacs2002}

\begin{equation}
SN_I=\frac{\delta_1}{\sigma}\Bigg(\frac{1}{\sqrt{N_{{\rm in}}}}+\frac{1}{\sqrt{N_{\rm out}}}\Bigg)
\label{eq:SNI}
\end{equation}

\noindent
where $\delta_1$ is the transit depth of the primary eclipse estimated by the difference between the average of the in-transit measurements and the average of the out-of-transit measurements. Such quantity is divided by the error (of the average's difference of the in and out of transit measurements), where we assumed uniform noise across the lightcurve, represented by the standard deviation of the out of transit measurements ($\sigma$).

\noindent
The effective signal to noise defined above is calculated (as most of the quantities we defined) on the lightcurve folded with the period corresponding to the highest peak in the BLS power spectrum. It is a measure of how well transit signals occurring at regular intervals of time identical to the chosen period and phase add up constructively.

\subsubsection{Effective S/N of the secondary eclipse}

To check for the presence of secondary eclipses the effective S/N of the secondary eclipse can be used

\begin{equation}
SN_{II}=\frac{|\delta_2|}{\sigma}\Bigg(\frac{1}{\sqrt{N_{{\rm in}}}}+\frac{1}{\sqrt{N_{\rm out}}}\Bigg)
\label{eq:SNII}
\end{equation}

\noindent
where $\delta_2$ is the estimated depth of the secondary eclipse. Such quantity is the difference between the average of the measurements within the secondary eclipse phase interval (located exactly at mid-phase between primary eclipses) and the out-of-secondary eclipse measurements. The time interval corresponding to the measurements within the secondary eclipse was
considered identical to the transit duration reported by the BLS. Since we are uninterested to discriminate between positive and negative flux variations at the secondary eclipse phase we took the absolute value of the eclipse depth in Eq.~\ref{eq:SNII}.

\subsubsection{Effective S/N of the tertiary eclipse}

Similarly to the previous two cases, the S/N of the tertiary eclipse can be define as follows

\begin{equation}
SN_{III}=\left\{\frac{|\delta_3|}{\sigma}\Bigg(\frac{1}{\sqrt{N_{{in}}}}+\frac{1}{\sqrt{N_{out}}}\Bigg)\right\}_{\rm MAX(0.2<\phi<0.8)}
\label{eq:SNIII}
\end{equation}

\noindent
By tertiary eclipses we intend here any eclipse occurring at phases possibly different from the primary and secondary phase. Eccentric eclipsing binaries can produce secondary eclipses which are not found exactly at mid-phase between primary eclipses, so it is useful to check for these signals.
The quantity $\delta_3$ is the depth of the tertiary eclipse and it is calculated as the maximum absolute difference of the flux measurements within and outside a sliding window (the same length of the transit duration given by the BLS) centered at orbital phases comprised between $\phi=$0.2 and $\phi=$0.8 (where the primary transit occurs at $\phi=$0).

\subsubsection{Signal Detection Efficiency (SDE)}

The SDE was defined in \cite{kovacs2002} and it is equal to

\begin{equation}
SDE=\frac{SR_{\rm peak}-<SR>}{sd(SR)}
\label{eq:SDE}
\end{equation}

\noindent
On the contrary of the previous quantities, this feature is extracted from the BLS power spectrum, where SR$\rm_{peak}$ is the power of the BLS peak, \textit{<SR>} the average power and \textit{sd(SR)} the standard deviation of the BLS power spectrum.

\subsubsection{Average Signal Detection Efficiency of aliasing peaks}

Periodic signals usually originate a cascade of aliases of the primary peak in the BLS power spectrum. To check for their presence we defined the average SDE over the nine peaks closest to the primary on both the low and high frequency intervals of the spectral window ($\rm N_{aliases}$)

\begin{equation}
SDE_{\rm AL}=\frac{\sum_{\rm i=1}^{\rm N_{aliases}}\frac{SR_{\rm i}-<SR>}{sd(SR)}} {\rm N_{aliases}}
\label{eq:SDEAL}
\end{equation}

\noindent
 This feature (as well as the SDE) is extracted from the BLS power spectrum, where SR$\rm_{i}$ is the power of the i-th BLS peak considered, \textit{<SR>} the average power and \textit{sd(SR)} the standard deviation of the BLS power spectrum.

\subsubsection{Effective S/N of odd/even transits }

As reported above, eclipsing binary signals may be revealed by calculating the odd ($\delta_{\rm odd}$) and even ($\delta_{\rm even}$) transit depths. A simple metric can be introduced to detect odd/even transit depth variations

\begin{equation}
SN_{\rm OE}=\frac{|\delta_{\rm odd}-\delta_{\rm even}|}{\sigma}
\Bigg(\frac{1}{\sqrt{N_{{\rm in,even}}}}+\frac{1}{\sqrt{N_{\rm in,odd}}}\Bigg)
\label{eq:SNOE}
\end{equation}

\noindent
where $N_{\rm in, even}$ and 
$N_{\rm in, odd}$ are the number of in-transit measurements in the even and in the odd transits, respectively.

\subsubsection{Out Of Transit (OOT) variability}

Another interesting morphological feature that can permit to identify likely eclipsing binary stars is the presence of phase-locked flux modulations with the same periodicity of the transiting body. Usually these flux variations can be produced by proximity effects (e.g. reflections, ellipsoidal modulations) between two close orbiting bodies and are expected to be much stronger for stellar than for planetary companions. To identify these flux modulations we performed an harmonic fit of the lightcurve by assuming the following model

\begin{equation}
f=A\cos(\omega t)+B\sin(\omega t)+C
\end{equation}

\noindent
where $f$ is the flux, $\omega=\frac{2\pi}{P}$, and $P$ is the period corresponding to the highest peak in the BLS power spectrum. We then calculated

\begin{equation}
R_{\rm OOT}^2=1-\frac{\sum_{\rm i=1}^{N_{\rm out}} (f_{\rm i}-[A\cos\omega t_{\rm i}+B\sin\omega t_{\rm i}+C])^2}{\sum (f_{\rm i}-C)^2}
\label{eq:R2OOT}
\end{equation}\noindent
where N$_{\rm out}$ is the number of out-of-transit measurements where the summatory is performed. R$\rm^2_{OOT}$ is a measure of how much variance is explained by the complete harmonic model with respect to the sole constant component of the model. We have that R$\rm^2_{OOT}$ < 1 and, the better the harmonic model fits the the data, the closer R$\rm^2_{OOT}$ is to unity. This
parameter was calculated both using the LC1 lightcurves and using the LC2 lightcurves and the maximum value was taken.

\subsubsection{Fractional transit duration}

The fractional transit duration ($q$) is equal to

\begin{equation}
q=\frac{\tau}{P}
\label{eq:q}
\end{equation}
\noindent
where $\tau$ is the transit duration and $P$ is the orbital period. 

\subsubsection{Point to point statistic inside and outside the transit}

A point-to-point noise estimate within and outside the transit window can help to identify potentially spurious transits. It is defined as 

\begin{equation}
P2P_{\rm IO}=\sqrt{\frac{\sum_{in} (f_{i}-f_{i-1})^2}{\sum_{out} (f_{i}-f_{i-1})^2}}
\label{eq:P2PIO}
\end{equation}

\noindent
where the numerator is evaluated on the in-transit measurements and the denominator on the out-of-transit measurements (in the phase folded lightcurve). 

\subsubsection{Simmetry of folded lightcurve}

A measure of the symmetry of the folded lightcurve is also a useful quantity to consider and it can be obtained with

\begin{equation}
P2P_{\rm S}=\sqrt{\frac{\sum_{i=1}^{N} (f_{i}-f_{i-1})_{|\phi|}^2}{\sum_{i=1}^{N} (f_{i}-f_{i-1})_{\phi}^2}}
\label{eq:P2PS}
\end{equation}

\noindent
where the numerator is evaluated after ordering the measurements as a function of the absolute value of the phase ($|\phi$|)\footnote{Assuming $\phi=0$ at the transit time and the phase running from $\phi$=-0.5 and $\phi$=0.5} and the denominator is evaluated on the nominal phase-folded lightcurve.

\subsubsection{Estimated transiting body radius}

The radius of the transiting body can be estimated from

\begin{equation}
r=\sqrt{\delta_1}R_{\star}
\label{eq:radius}
\end{equation}\noindent
where $\delta_1$ is the transit depth of the primary eclipse and $R_{\star}$ is the stellar radius.  

\noindent
As it can be noted, the transiting object radius is the only feature which has a physical dimension in our set. The reason to include it stems from the fact that the stellar radius is now well constrained by {\it Gaia} and thus the transit depth can be effectively converted to transiting object radius for all the targets. The gain is huge, since  without such information many transiting candidates could be erroneously considered plausible or erroneously considered implausible due to the larger ambiguity inherent to the use of the sole transit depth.

\subsection{Simulations}
\label{sec:simulations}
To train the model to recognize transiting planets, we used a set of simulations. From the pool of stars we used to construct eigenvectors described in Sect.~\ref{sec:cotrending}, we randomly selected  a sample of 20000 stars taking care of the fact that the sample selected reproduced the global properties of the whole sample of stars we analyzed. By assuming the \cite{mandel2002} model we first injected a set of transiting planets (one per star) with radii uniformly distributed between one Earth radius and 2.5 Jupiter radii. The orbital period was uniformly chosen between 0.25 days and the total time spanned from the first to the last measurement of each lightcurve. The transits were injected in the raw lightcurves (Sect.~\ref{sec:photometry}) and then all the post-correction procedure described in the previous sections was applied. The lightcurves were searched for transits using the BLS (Sect.~\ref{sec:transit_detection}). We then isolated the lightcurves for which the injected planets were recovered. This was done by looking at the period corresponding to the highest peak in the BLS periodogram and selecting those lightcurves for which the relative difference between the recovered period and the injected period was smaller than 1\%. 

\noindent
A similar procedure was applied to generate false positives. We first considered  detached eclipsing binaries with equal mass ratio, radius of the secondary uniformly chosen between 2.5 R$\rm_J$ and the primary radius, secondary to primary eclipse ratios comprised between 0.1 and 0.9, orbital periods spanning the same temporal range as in the planet case and circular orbits. The eclipsing binary signal was injected into the same pool of constant stars used for the planet sample, and the same procedure explained above was applied to isolate
the sample of recovered eclipsing binaries. In this case we considered both the lightcurves for which the relative difference between the recovered period and the injected period was less than 1\% and those for which the relative difference between twice the recovered period and the simulated period was less than 1\%. Once primary and secondary eclipse depths are very similar the BLS tends to recover half of the correct period. These stars are useful to train the odd/even metric, while those for which the recovered period is consistent with the simulated one are useful to train the secondary eclipse metric. 

\noindent
A different set of eclipsing binary simulations was performed, but this time the secondary eclipse was arbitrary shifted of at most a quarter of the orbital period before or after the nominal secondary eclipse instant. This procedure was adopted to account for eccentric binaries which can be photometrically recognized precisely from the shift of the secondary relative to the primary eclipses.

\noindent
A final set was prepared to simulate purely rotationally modulated variables (with perfectly sinusoidal shape) with very short orbital periods (comprised between 0.25 days and 1 day) and amplitudes between 0.05 and 0.3 magnitudes.

\noindent
In total we therefore constructed six different samples: constant stars, planets, eclipsing binaries (nominal secondary eclipse timing), eclipsing binaries (with recovered period equal to half the injected one, odd/even variables), 
eclipsing binaries (with shifted secondary eclipses) and sinusoidal ultra-short variables. For simplicity, in the rest of this discussion the eclipsing binaries and the ultra-short variables are referred globally as variables.

\noindent
To build a final global training set of simulated objects, we regrouped in different proportions the various categories reported above. 
In sample 1, we considered 50\% of constant stars plus all variables (with equal proportions of the constant and all variables' subcategories) plus a 50\% of planets. In sample 2, we considered a 33\% proportion of constant stars, 33\% of all kinds of variables, 33\% of planets. In sample 3, we considered 50\% of planets and 25\% of constant stars and the remaining 25\% equally subdivided among all variables' categories. These training dataset were built separately for the lightcurves derived with  aperture 1 and for the lightcurves derived with aperture 2 for which the {\it TESS} magnitude of the simulated stars was limited to {\it T}$<$11. The total dimension of the three samples was equal to 16181, 16175 and 16344 for aperture 1 and it was equal to 15893, 15981 and 15951
for aperture 2.
We also build six corresponding testing datasets (one for each combination of samples and photometries). The composition of these test samples was equal to 50\% planets and 50\% all the remaining categories (in equal proportion). The dimension of the testing samples was equal to 4000 stars.

\noindent
In all cases the problem was treated as a binary classification problem, that is we did not distinguished between variables and constant stars attributing to them the negative class, while we  attributed the positive class to the simulated planets. Below we compared the performances of the RF algorithm when trained on these different samples.

\subsection{Training}

\begin{table}
	\centering
	\caption{Performance measures of the RFC for the three samples described in the text
	and photometric aperture equal to 1 pix. FPR=False Positive Rate; TPR=True Positive Rate, FNR=False Negative Rate, TNR=True Negative Rate, AUROC=Area Under the ROC curve.}
	\label{tab:performance_RFC_ap1}
	\begin{tabular}{cccccc} 
		\hline
		 Sample & FPR(\%) & TPR(\%) & FNR(\%) & TNR(\%) & AUROC \\ 
		\hline
         1 & 1.0 & 89.6 & 10.4 & 99.0 &  0.9956 \\ 
         2 & 1.0 & 90.4 & 9.6 & 99.0 &  0.9937 \\ 
         3 & 1.0 & 92.0 & 8.0 & 99.0 &  0.9936 \\ 
		\hline
	\end{tabular}
\end{table}

\begin{table}
	\centering
	\caption{Performance measures of the RFC for the three samples described in the text
	and photometric aperture equal to 2 pix. FPR=False Positive Rate; TPR=True Positive Rate, FNR=False Negative Rate, TNR=True Negative Rate, AUROC=Area Under the ROC curve.}
	\label{tab:performance_RFC_ap2}
	\begin{tabular}{cccccc} 
		\hline
		 Sample & FPR(\%) & TPR(\%) & FNR(\%) & TNR(\%) & AUROC \\ 
		\hline
		 1 & 1.0 & 94.3 & 5.7 & 99.0 & 0.9968 \\ 
		 2 & 1.0 & 94.8 & 5.2 & 99.0 & 0.9968 \\ 
		 3 & 1.0 & 95.4 & 4.6 & 99.0 & 0.9962 \\ 
		\hline
	\end{tabular}
\end{table}

\begin{figure}
\includegraphics[width=0.95\columnwidth]{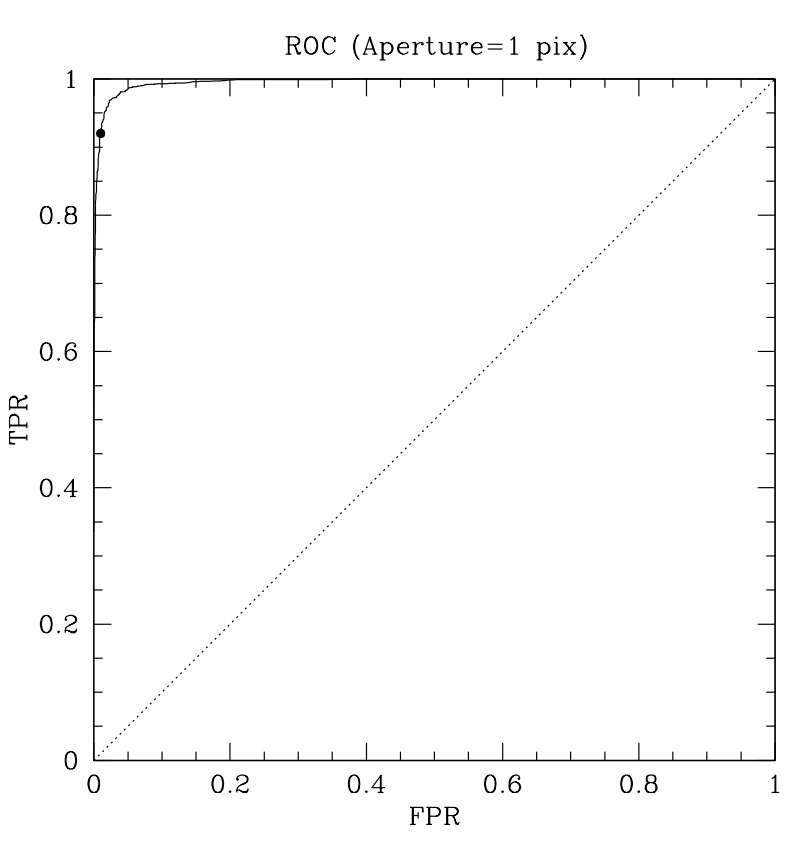}
\includegraphics[width=0.95\columnwidth]{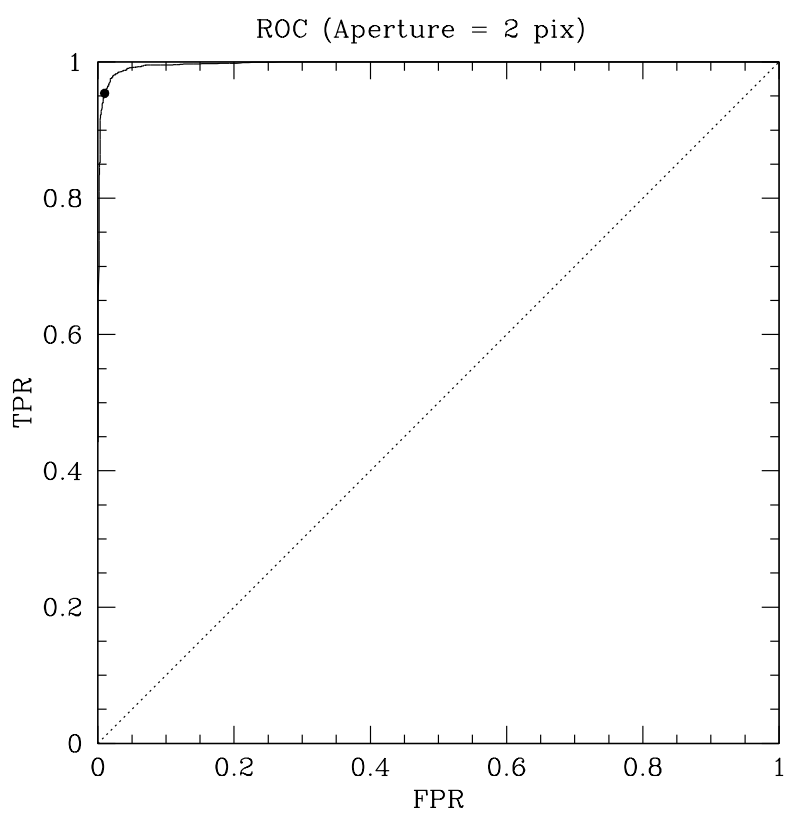}
\caption{The solid curves denote the ROC curves for the best models of aperture 1 (top) and aperture 2 (bottom). The black points indicate the FPR and TPR corresponding to the adopted detection thresholds, while the dotted lines represent the perfectly random classifier.
}
\label{fig:ROC}
\end{figure}

We then train the Random Forest model using the Caret package {\it train} function in R,
using a 10-fold cross validation method with 5 repeats. This approach first randomly shuffles the data, then creates ten partitions, nine of which are used to train the model and one is used for testing. Each partition is hold out one time for testing and the others used for training the model and a score is attributed to each one of these combinations of testing/training subsets. The entire procedure is repeated five times with a different random shuffling and partitioning of the data. Scores are then averaged to evaluate the model performances. We repeated the above procedure by tuning the {\it mtry} parameter of the RF on a grid of integer values comprised between 1 and 6, while the number of trees ({\it ntree}) was hold fixed at its default value of 500 trees. The {\it mtry} parameter controls the number of predictors randomly chosen to split each node in a tree. The Area Under the Receiving Operating Characteristics curve (AUROC) was chosen as the metric to evaluate the model performaces for each value of {\it mtry}. In all cases the result was that the best value for {\it mtry} was either {\it mtry}=3 or {\it mtry}=4. 
The AUROC values corresponding to the best models of each sample and aperture are reported in Table~\ref{tab:performance_RFC_ap1} (aperture 1) and Table~\ref{tab:performance_RFC_ap2} (aperture 2). To choose the best model to adopt for each aperture, we decided to fix the False Positive Rate (FPR) at a value of 1$\%$, and selected the model providing the largest value of the True Positive Rate (TPR). Choosing a low false positive threshold is crucial in this kind of experiment, in order to avoid an overwhelmingly large number of false positive candidates with respect to the true positive candidates. As a comparison, \cite{barclay2018}
estimated an hit rate (the ratio of planets detected to observed stars)$<0.75$\%, for stars in the FFIs. By looking at the results in Table~\ref{tab:performance_RFC_ap1} and Table~\ref{tab:performance_RFC_ap2} we decided to adopt the model relative to sample 3 for both apertures. The corresponding probability thresholds of the RFC are equal to P$_{\rm RCF_1}$=0.7401 and P$_{\rm RCF_2}$=0.6421 for aperture 1 and aperture 2, respectively. 
The ROC curves we obtained for the best models of each aperture are shown in Fig.~\ref{fig:ROC}. The continuous lines represent the TPR vs FPR for the best models, while the dotted lines along the diagonals represent the perfectly random classifier. The more a classifier is able to discriminate between the two classes the more the curve should be closer to the top left corner of the diagram (which indicates the perfect classifier). The black dots in
Fig.~\ref{fig:ROC}, visualize on the ROC curves the TPR and FPR corresponding to our adopted detection thresholds.

\subsection{Variable Importance}

\begin{figure}
\includegraphics[width=0.98\columnwidth]{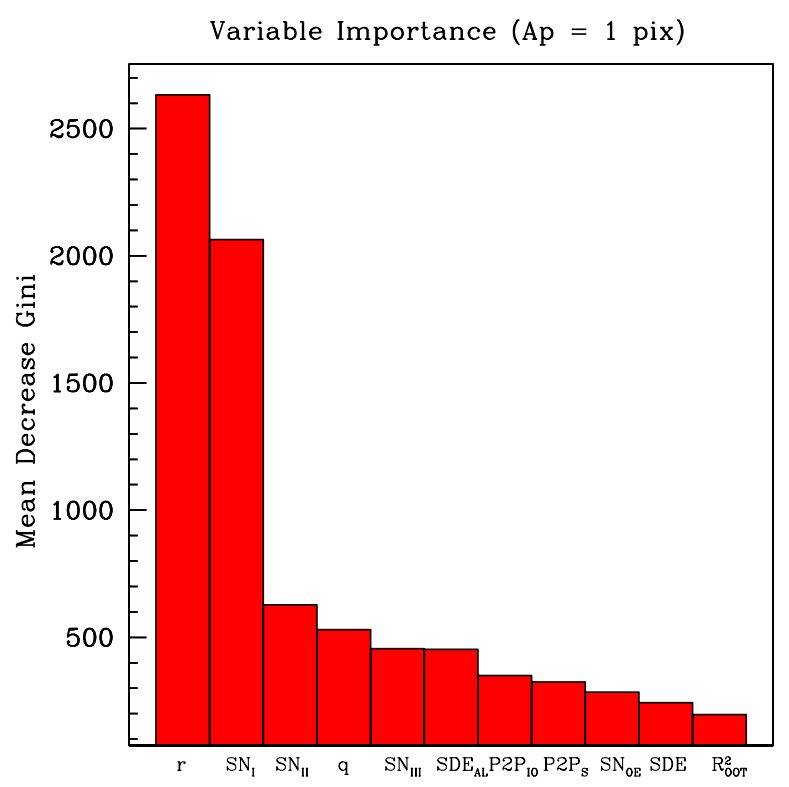}
\includegraphics[width=\columnwidth]{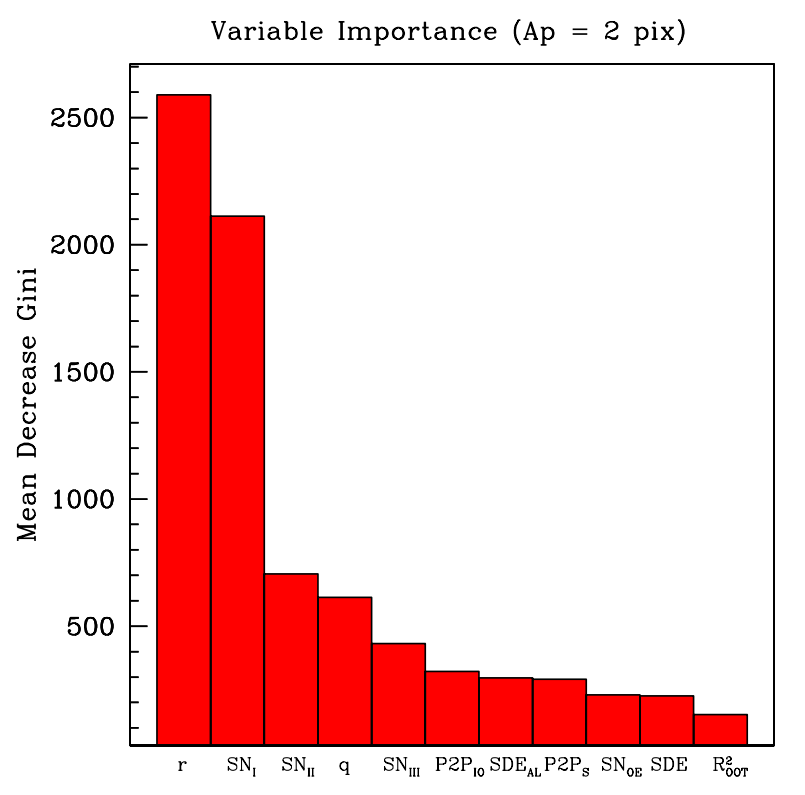}
\caption{Mean decrease in the Gini impurity index for each variable involved in the classification
process. The higher the decrease in the Gini Impurity, the more the variable is important.
}
\label{fig:importance}
\end{figure}

In Fig.~\ref{fig:importance}, we represent the importance that each variable has in the classification process in terms of the Mean Decrease Gini it produces. Such metric quantifies the total decrease in node impurity weighted by the proportion of samples reaching that node, averaged over all trees in the forest. The higher the Mean Decrease Gini, the more important is the variable. From Fig.~\ref{fig:importance} it results that for both apertures, the radius of the transiting object ($r$) and the signal to noise of the primary eclipse ($SN\rm_{I}$) stand out in their importance with respect to the other variables. Also interesting to note is the fact that the signal detection efficiency metric ($SDE$) appears in general less important in the classification process that the $SDE\rm_{AL}$ which quantifies the power in the periodogram of aliases of the primary peak. Evaluating the power split on the moltitude of alias peaks appears more important for the classification process than evaluating the power of the sole primary peak.

\subsection{Performances}

The procedure described in the previous Section permits to determine the performances of the RFC algorithm to disentangle plausible transiting planetary candidates from other kind of false positive events. It is important however to establish also the overall performance of the transit search algorithm which should also account for the performance of the BLS algorithm to correctly recover transit events. By using the same sample of stars we previously employed, we simulated a sample of 35000 transiting planets with radii randomly selected between 1 R$_{\oplus}$ and 25 R$_{\oplus}$ and periods randomly spanning the range between 0.25 days and the interval of time between the first and last observation of each lightcurve. Transits were injected in the raw lightcurves and we then repeated the full post-correction analysis, the BLS search and applied the RFC algorithm. If the absolute relative difference between the BLS recovered period and the original injected period was smaller than 1$\%$ and the probability returned by the RFC was larger than the adopted detection threshold we considered the injected planet as recovered. The transit detection efficiency was then defined as the ratio of the number of recovered planets to the number of simulated planets. We analyzed the results in a bidimensional grid presenting planetary radii against periods subdivided in steps sizes of 1 R$_{\oplus}$
and 1 day, respectively.

\begin{figure}
\includegraphics[width=0.98\columnwidth]{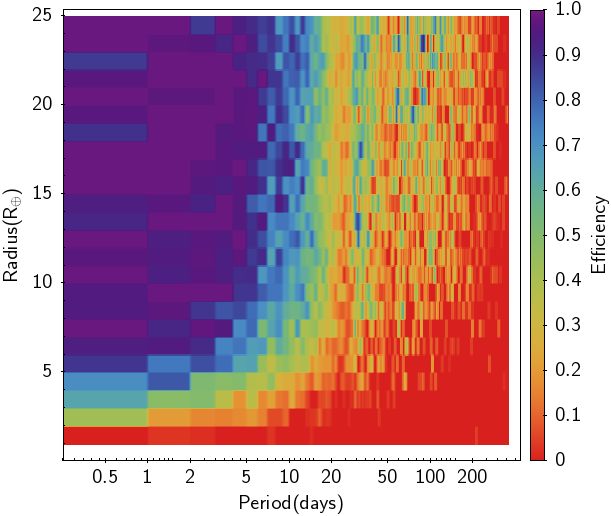}
\includegraphics[width=\columnwidth]{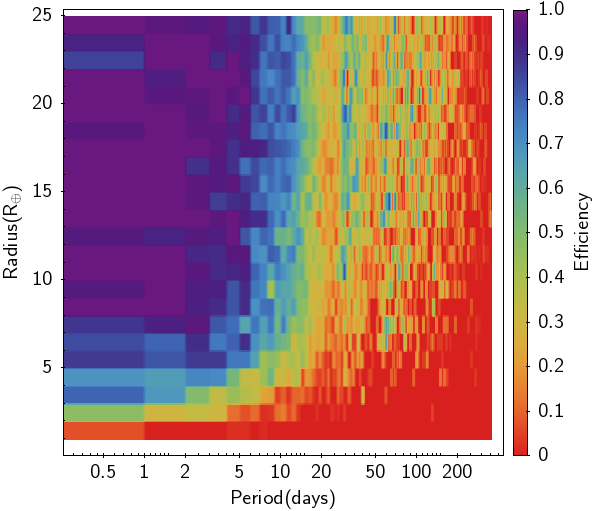}
\caption{Transiting planets' detection efficiency as a function of injected planet radius and orbital period for the sample of stars analyzed with photometric aperture equal to 1 pix (top) and equal to 2 pix (bottom).
}
\label{fig:efficiency}
\end{figure}

\noindent
The results are presented in Fig.~\ref{fig:efficiency} for the sample of stars analyzed with aperture equal to 1 pix (top) and equal to 2 pix (bottom). In general the detection efficiency appears similar for both samples. It quickly drops below 50\% for planets with radii below 3 $R_{\oplus}$ in the short period planets domain and for orbital periods P>20 days.

\section{Centroid motion}
\label{sec:centroid_motion}

As it is well known, one of the issues of transit search programs is that they usually employ detectors with large pixel scales in order to monitor large
field of views and measure the brightness of many stellar sources. In this way, it is expected that several stars may lie on the same 
photometric aperture especially in the most crowded fields, and consequently the signal of background eclipsing binaries can be diluted with the light of 
a brighter target star mimicking shallower planetary transit signals. One of the methods used to recognize these false positive signals is to monitor the center of light motion of a given target and judge if any correlated motion occurs during a transit event pointing away from the target. Here we used the
difference images created during data reduction (Sect.~\ref{sec:differential_image_photometry}) to study centroid shifts. For each one of these images
we extracted the flux weighted centroids on four concentric circular apertures centered on the targets. The circular apertures' radii were set equal to 1 pix, 2 pix, 3 pix
and 4 pix, respectively. The contribution of each pixel to the weighted mean centroid of a given aperture was set equal to the absolute pixel flux multiplied by the area of the pixel contained within the aperture. The centroids measurements were registered on the same astrometric reference 
system of the reference astrometric image of a given sector, camera and CCD (Sect.~\ref{sec:differential_image_photometry}) and converted to RA and DEC using the 
corresponding WCS information reported on the image header. Multi-sector observations were merged together after conversion to sky coordinates.
Since the fluxes of both the target and contaminant stars change over time even outside transit windows,  the centroid position based on the flux weighted measurements typically varies over time as well. This low frequency variations can be modeled out and corrected for. By using only out of transit centroid measurements we repeated the splining procedure reported in Sect.~\ref{sec:normalization} (applying it to centroid measurements). We therefore calculated the shift both in right ascension and in declination between the spline model and the median centroid and corrected all measurements to the median centroid estimate. After that, we considered only in-transit measurements and performed a PCA decomposition of the centroid shifts measurements calculating the two principal components and the standard deviations along these components. This procedure was repeated for each aperture separately. The centroid measurements, the orientation of the principal components' axes and the associated standard deviations can be used to define four bivariate Gaussian distributions (one for each aperture).
We then considered all sources within a radius of 3 arcmin from each target drawn from the {\it Gaia} catalog
and sorted them as a function of their Mahalanobis distance
\citep{mahalanobis1936} from the centroid of each aperture distribution. The source with the smallest distance was elected as the most likely source associated with the observed distribution of each aperture. If the elected source was the target star we attributed a rank$=$1 otherwise a rank$=$0. Therefore, a criterium that can be adopted to consider a given target as the source of variability is to require all ranks associated with the four different apertures to be equal to 1. Hereafter we will refer to these calculated ranks as the Mahalonobis ranks.

\begin{figure}
\includegraphics[width=\columnwidth]{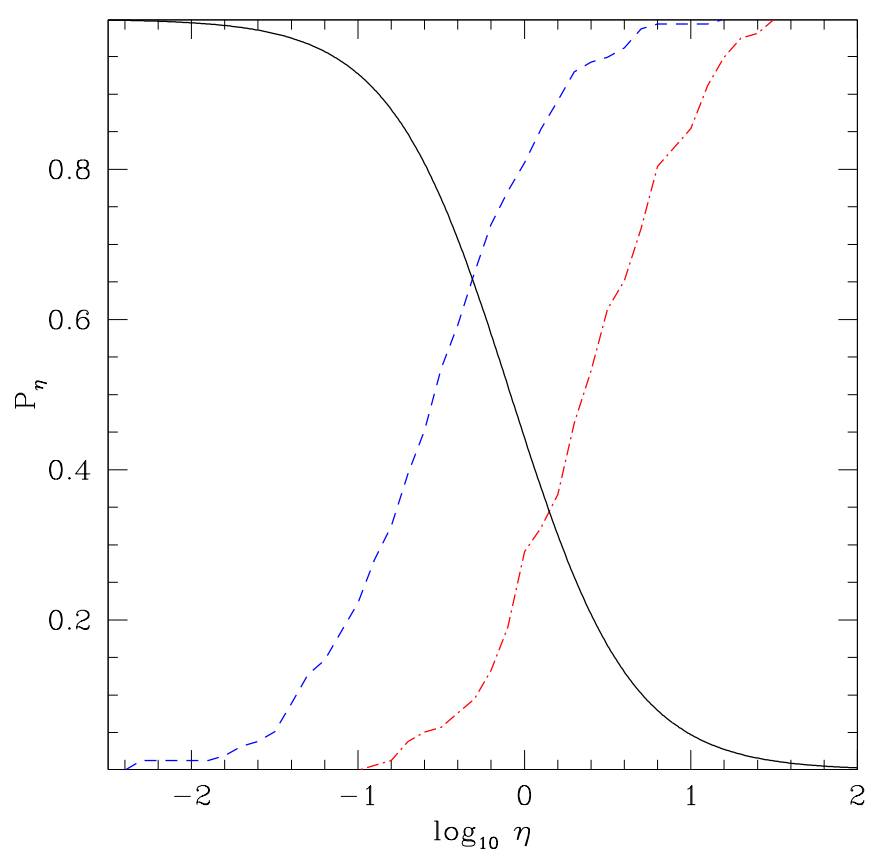}
\caption{Cumulative distributions of the average local density metric (log$_{10}\,\eta$) for targets identified as the
correct source of variability (blue dashed line)
and for misidentified sources of variability (red dashed-dot line) along with the best fit logistic model (black line) representing
the probability of correct source identification (P$\rm_{\eta}$) as a function of log$_{10}\,\eta$ (as defined in Eq.~\ref{eq:peta}).}
\label{fig:probeta}
\end{figure}

\noindent
In practice, the effectiveness of the above procedure depends also on different factors. For example, we can expect that the larger the local number
density of stars, the more likely is that a given source will fall by chance close enough to the centroid distributions that it could be erroneously identified as the source of variability. To study this problem more in detail, we analyzed a random sample of known eclipsing binary sources and known transiting planets retrieved from the International Variable Star Index database (VSX). We derived the centroid distributions for each one of them and analyzed the cases where the selected variable was considered as the source of variability accordingly with the procedure described above (that is all Mahalanobis ranks were equal to 1), and the cases in which it was not. In our final list of 316 stars, half of them belonged to the first category and the other half to the second category.
For each variable, we also registered the local average number density of stars 
($\eta$) obtained as the average of the number of stars per {\it TESS} pixel measured in each one of the four apertures we considered. All the stars we considered in the calculation were retrieved from the {\it Gaia}  DR2 catalog. The density of the considered objects was ranging from approximately 0.01 stars px$^{-2}$ to 100 stars px$^{-2}$.
The result is shown in Fig.~\ref{fig:probeta}, where the cumulative distribution of the correctly identified sources is represented in blue and the cumulative distribution of the misidentified sources is represented in red. Both distributions are presented as a function of the $\log_{10}\,\eta$.  Objects in the first group were attributed to Class=1 and objects in the second group to Class=0. The Class variable is then a binary dependent variable which can be statistically modeled to provide the probability that an object belongs to one class as a function of the predictor variable $\log_{10}\,\eta$. To this purpose we can use a logistic function, for which the log-odds \textit{l} of a star to belong to Class=1 is a linear combination of the predictor variable

\begin{equation}
\label{eq:logodds}
   l=ln\frac{1}{1-{\rm P_\eta}} = \alpha_1+\beta_2\,\log_{10}\,{\eta}
\end{equation}

\noindent
where $\rm P_{\eta}$ is the probability that Class=1

\begin{equation}
\label{eq:peta}
   {\rm P_{\eta}}=\frac{1}{1+e^{-(\alpha_1+\beta_1\,\log_{10}\,{\eta})}}
\end{equation}

\noindent
the coefficients $\alpha_1$ and $\beta_1$ are obtained by iterative non-linear optimization techniques. The best fit coefficients we found are reported in Table~\ref{tab:probeta} and the best fit logistic function is represented by the continuous black line in Fig.~\ref{fig:probeta}.
The result we obtained indicates that for a local average number density equal to 0.84 stars pix$^{-2}$ the probability of correct source identification is 50$\%$. Using this model, a threshold can be conveniently chosen for the local number stellar density in order to increase the chances of correctly identify a source of variability.

\begin{figure*}
\includegraphics[width=\columnwidth]{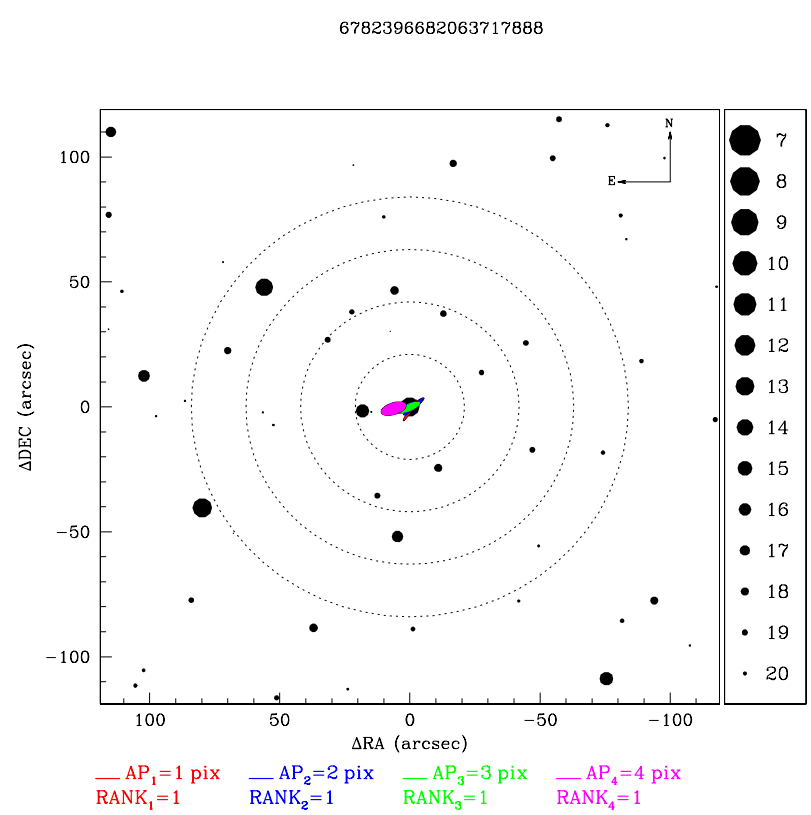}
\includegraphics[width=0.975\columnwidth]{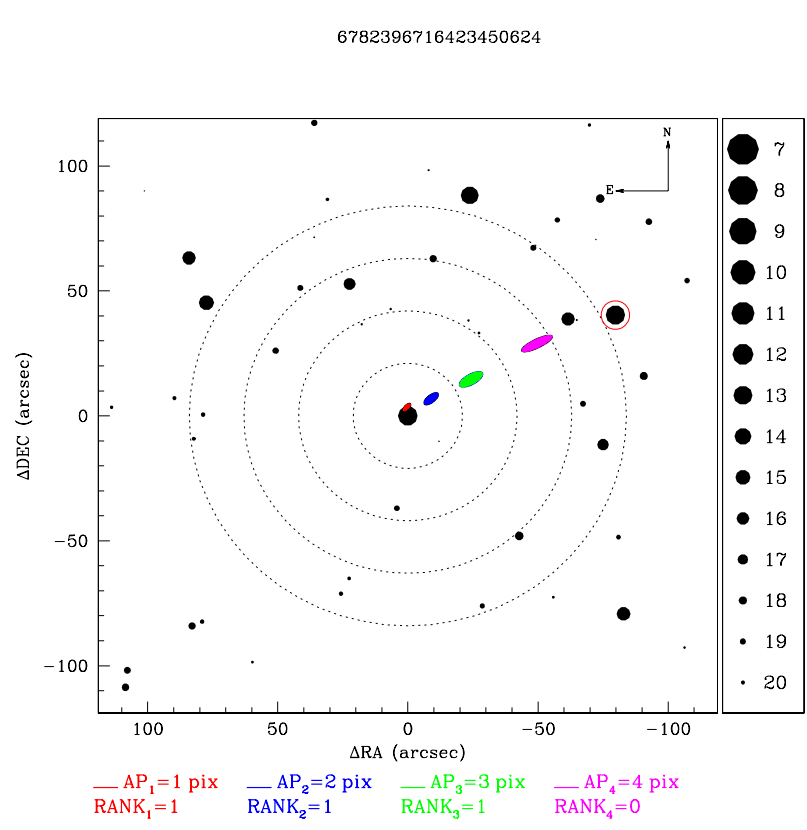}
\caption{Two examples of application of the centroid algorithm. The plot on the left is centered on
the star 6782396682063717888 ({\it Gaia} DR2 ID), a known eclipsing binary. 
The colored ellipses represent the position and dispersion of the centroid metric measurements relative to this target for the four concentric apertures discussed in the text (and represented by the four dotted circles). The probability of source association P$\rm_{D}$ calculated from Eq.~\ref{eq:pd} is equal to 91$\%$. On the bottom, the Mahalanobis distance ranks are also reported. On the right, the same measurements are repeated for the close-by star 6782396716423450624,  South-East of the eclipsing binary (highlighted in this panel with the red circle). The probability of source association P$\rm_{D}$ is equal to 20$\%$, in this case. The legend on the right of each plot is relative to the stars' magnitudes in the {\it Gaia} {\it G} band. All stars represented in the figure are taken from the {\it Gaia} DR2 archive.
}
\label{fig:targ_cont}
\end{figure*}

To further strengthen variable identification,
other information can be exploited. For example, we should expect that if a variable source corresponds with the target (which is by definition at the center of the apertures), then the centroid drift from the center should be small and the centroid measurements should be clustered around the target with possibly small dispersions. If instead the source of variability lies apart from the target we could expect the centroid mesurement to be shifted in the direction of that source and the the principal components' axes oriented correspondingly. 
Fig.~\ref{fig:targ_cont} presents an example.
On the left panel the apertures (denoted by the four concentric circles) are centered on a known
eclipsing binary star (star  6782396682063717888 according to {\it Gaia} DR2). The colored ellipses represent the centroid motion distributions for the four different apertures. 
In this case all centroid measurements are clustered around the target which is effectively the source of variability. The Mahalanobis ranks are as well reported on the bottom of the figure and they are all equal to 1. On the right panel the apertures are centered on a close-by star 
(star 67823967164234506) on the South-East of the eclipsing binary. In this case the centroid distribution measurements progressively depart from the target, precisely in the direction of the eclipsing binary and the outermost Mahalonobis rank is equal to 0.

It is possible to exploit this behaviour of the centroid motion measurements to construct a metric that will provide the probability of association with a given target. In this case we analyzed a sample of known variable sources and surrounding stars. For each one of them we calculated  the following quantity:

\begin{equation}
\label{eq:centroid}
    \log_{10} {\rm D}=\frac{1}{2}\log_{10}{\Big(\overline{r}^2+\overline{\sigma}^2\Big)}
\end{equation}

\noindent
where

\noindent
\begin{equation*}
    \overline{r}=\frac{\sum_{i=1}^{i=4}\,w_i\,r_i}{\sum_{i=1}^{i=4}\,w_i}
\end{equation*}

\noindent
\begin{equation*}
    \overline{\sigma}=\frac{1}{\sqrt{\sum_{i=1}^{i=4}\,w_i}}
\end{equation*}

\noindent
\begin{equation*}
    w_i=\frac{1}{\sigma^2_{1,i}+\sigma^2_{2,i}}
\end{equation*}

\noindent
with $\overline{r}$ the weighted average of the apertures' flux weighted centroids ($r_i$), $\overline{\sigma}$ the corresponding weighted average error and $\sigma_{1,i}$,  $\sigma_{2,i}$ the principal components' standard deviations of each aperture probability density distribution.

\begin{figure}
\includegraphics[width=\columnwidth]{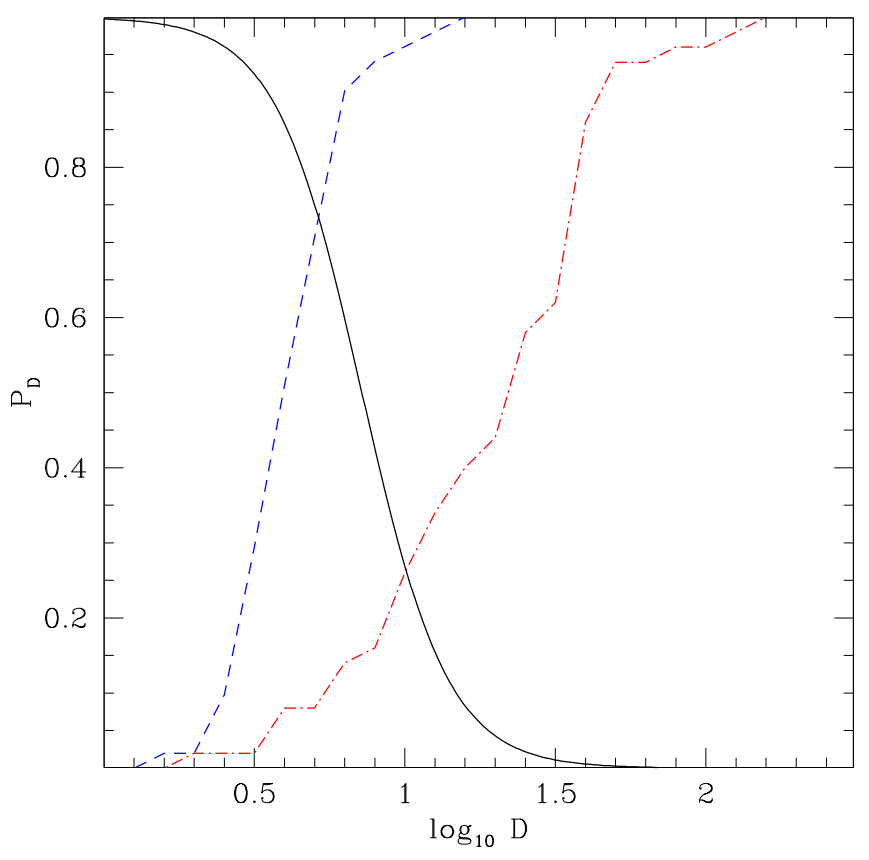}
\caption{Cumulative distributions of the centroid metric (log$_{10}\,$D) for transit events associated with the target source (blue dashed line)
and associated with surrounding sources (red dashed-dot line) along with the best fit logistic model (black line) representing the
probability of correct source identification (P$\rm_{D}$) as a function of log$_{10}\,$D
(as defined in Eq.~\ref{eq:pd}).}
\label{fig:probcentroid}
\end{figure}

\begin{table}
	\centering
	\caption{Best fit parameters of the logistic regression model related to the centroid metric log$_{10}$ $\eta$.}
	\label{tab:probeta}
	\begin{tabular}{cc} 
		\hline
		$\alpha_1$ & $\beta_1$ \\
		\hline
		 -0.2$\pm$0.1 &  -2.8$\pm$0.3 \\
		\hline
	\end{tabular}
\end{table}

\begin{table}
	\centering
	\caption{Best fit parameters of the logistic regression model related to the centroid metric log$_{10}$ D.}
	\label{tab:probcentroid}
	\begin{tabular}{cc} 
		\hline
		$\alpha_2$ & $\beta_2$ \\
		\hline
		 6.$\pm$1 & -7$\pm$1 \\
		\hline
	\end{tabular}
\end{table}

\noindent
In this case our sample consisted of 50 stars known to be eclipsing binaries or planets and 50 close-by surrounding stars randomly chosen in a region between 3 arcsec and about 3 arcmin from the binaries.
The distributions of the $\log_{10}\,\rm D$ metric for these two objects' categories are represented by the histograms shown in Fig.~\ref{fig:probcentroid}. Also in this case we can model the probability that an object is correctly associated with the known variables with a logistic function using this time as predictor variable $\log_{10}\,\rm D$
\begin{equation}
\label{eq:logodds_pd}
   l=ln\frac{1}{1-{\rm P_D}} = \alpha_2+\beta_2\,\log_{10}\,{\rm D}
\end{equation}

\noindent
where $\rm P_D$ is the probability of correct association

\begin{equation}
\label{eq:pd}
   {\rm P_D}=\frac{1}{1+e^{-(\alpha_2+\beta_2\,\log_{10}\,{\rm D})}}
\end{equation}

\noindent
the best fit coefficients $\alpha_2$ and $\beta_2$ are reported in Table~\ref{tab:probcentroid} and the best fit logistic function is represented by the continuous black line in Fig.~\ref{fig:probcentroid}.

\noindent 
The $\log_{10}\,\rm D$ metric combines both the centroid disposition and the its errors into a single quantity, averaged over all apertures. The result of the analysis can be interpreted as the fact that when the combined effect of the centroid disposition and its uncertainty amount to about 7.2 arcsec (that is $\sim34\%$  of a TESS pixel dimension) the probability that the target source is the source of variability is 50$\%$. This model can be used to quantify the probability (P$\rm_D$) that a given target is associated with the observed centroid motion distributions (and therefore with the observed transit events). For the situation represented in Fig~\ref{fig:targ_cont}, the eclipsing binary on the left panel has a probability of association equal to 91$\%$, whereas the close-by star on the right panel has a probability equal to 20$\%$.

\subsection{Dynamical constraints}
\label{sec:dynamical_constraints}

Additional information on the candidate planet's host stars can be obtained directly from the \textit{Gaia} DR2 catalog. In particular both radial velocities and their errors, as well as information on the quality of the astrometric solution, can be exploited to flag stars that potentially have stellar companions.

\noindent
We retrieved from the exoplanet orbit database \citep{han2014} a sample of known planet host stars for which the radial velocity semi-amplitude was determined and cross-matched it with the \textit{Gaia} DR2 catalog, requiring \textit{Gaia} radial velocities and their uncertainties to be defined. The sample consists of 746 planet hosts.

\noindent
The standard deviation ($\sigma_{V^{t}_{R}}$) of the \textit{Gaia} radial velocity measurements can be obtained inverting Eq.~1 of \cite{katz2019} which gives

\begin{equation}
\sigma_{V^{t}_{R}}= \sqrt{\frac{2N}{\pi}\big(\epsilon_{V_R}^2-0.11^2\big)}    
\end{equation}

\noindent
where $\epsilon_{V_R}$ is the radial velocity uncertainty and $N$ the number of eligible transits\footnote{The term \textit{transit} used in the Gaia documentation is synonymous of visit.} used to derive the median radial velocity, both of them reported in the \textit{Gaia} DR2 catalog.

\begin{figure}
\includegraphics[width=\columnwidth]{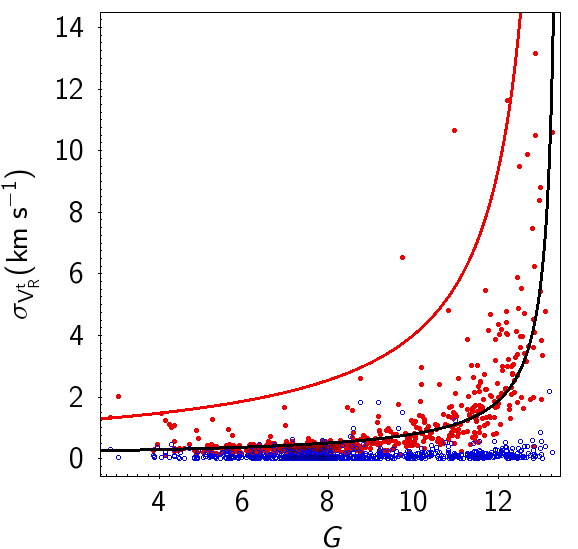}
\caption{
Standard deviation of the \textit{Gaia} DR2 radial velocity measurements ($\sigma_{V^t_R}$, red dots) versus
the \textit{Gaia} \textit{G} band magnitude for a sample of 746 known planet hosts stars.
Blue open circles show the radial velocity semi-amplitudes
measured with high-precision spectrographs. The black line denotes the best fit of the standard deviation measurements
and the red line the 5-$\sigma$ limit adopted in this work to flag suspected radial velocity variables.
}
\label{fig:sigma_RV}
\end{figure}

\noindent
In Fig.~\ref{fig:sigma_RV}, we show the \textit{Gaia} radial velocity standard deviation versus the \textit{G} band magnitude of the planet host stars (red points). Open blue circles show the radial velocity semi-amplitudes associated with the planets known to orbit around these stars. It is apparent that the \textit{Gaia} radial velocities are not precise enough to permit the detection of such planets. The radial velocity standard deviations shown in Fig~\ref{fig:sigma_RV} simply reflect the limiting precision of 
\textit{Gaia} measurements \citep{soubiran2018, katz2019}.
For most of the stars the radial velocity standard deviation is smaller than about 1 km$\rm\,s^{-1}$ down to \textit{G}$\simeq$10. After that, the \textit{Gaia} radial velocity precision quickly deteriorates. 

\noindent
We fit the $\sigma_{V^{t}_{R}}$ as a function of the \textit{G} band magnitude adopting an hyperbolic relation of the form

\begin{equation}
f(G)=\frac{A}{13.5-\textit{G}}
\end{equation}

\noindent
valid for \textit{G}$<$13.5 where the best fit coefficient \textit{A} is equal to \textit{A}=2.808. The black line in Fig.~\ref{fig:sigma_RV} represents this Equation, which is also our best fit model for the \textit{Gaia} radial velocity precision. It is important to recall that the \textit{Gaia} radial velocity precision depends on several factors beyond the apparent magnitude \citep{katz2019}, such as the effective temperature for example, which implies that such precision cannot be in general described by a unique relationship as we did above. However we are here only interested to set a conservative upper limit to flag potential radial velocity variability
and therefore we adopted a 5-$\sigma$
threshold which corresponds to the following Equation

\begin{equation}
5f(G)=\frac{14.04}{13.5-\textit{G}}
\end{equation}

\noindent
and it is represented by the red line in Fig.~\ref{fig:sigma_RV}. Stars for which the \textit{Gaia} $\sigma_{V^{t}_{R}}>5f(g)$ were flagged as potential binaries.

\noindent
Similar considerations can be done also for what concerns the astrometric signal. Sources which are unresolved or barely resolved in \textit{Gaia} DR2 may have poor astrometric solutions. Here we follow the approach described in \cite{evans2018a} who used the  Astrometric Goodness of Fit in the Along-Scan direction (gofAl) and the Significance of the Astrometric Excess Noise (astroExcessSig) as indicators of poorly-resolved binaries. Accordingly to \cite{evans2018a} confirmed binaries have astroExcessSig$>$5 and gofAL$>$20 and we adopted the same thresholds in this work.

\noindent
Both radial velocities and astrometric information should not be considered necessarily as conclusive indications that a given planetary candidate is a false positive. That is because the astrometric or radial velocity signals may be not associated with the observed transit events. They could also indicate the presence of additional (likely stellar) companions in the system. In our analysis we decided to report candidates for which either the radial velocity or the astrometric conditions defined above are satisfied and to flag them as suspected binaries.

\begin{figure}
\includegraphics[width=0.488\columnwidth]{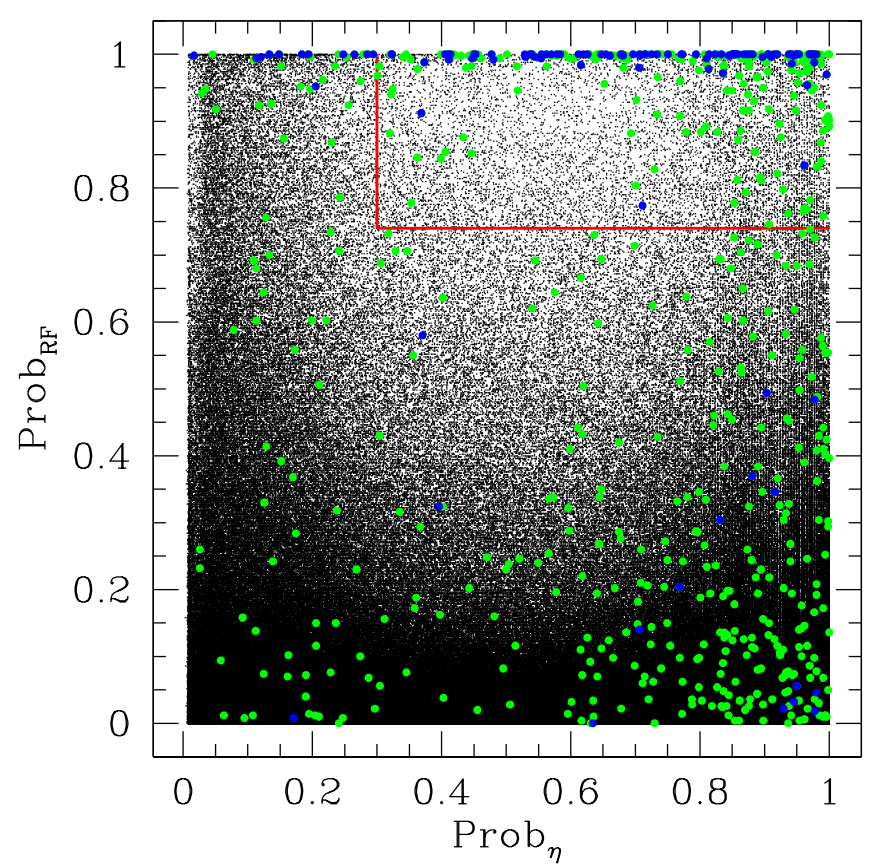}
\includegraphics[width=0.49\columnwidth]{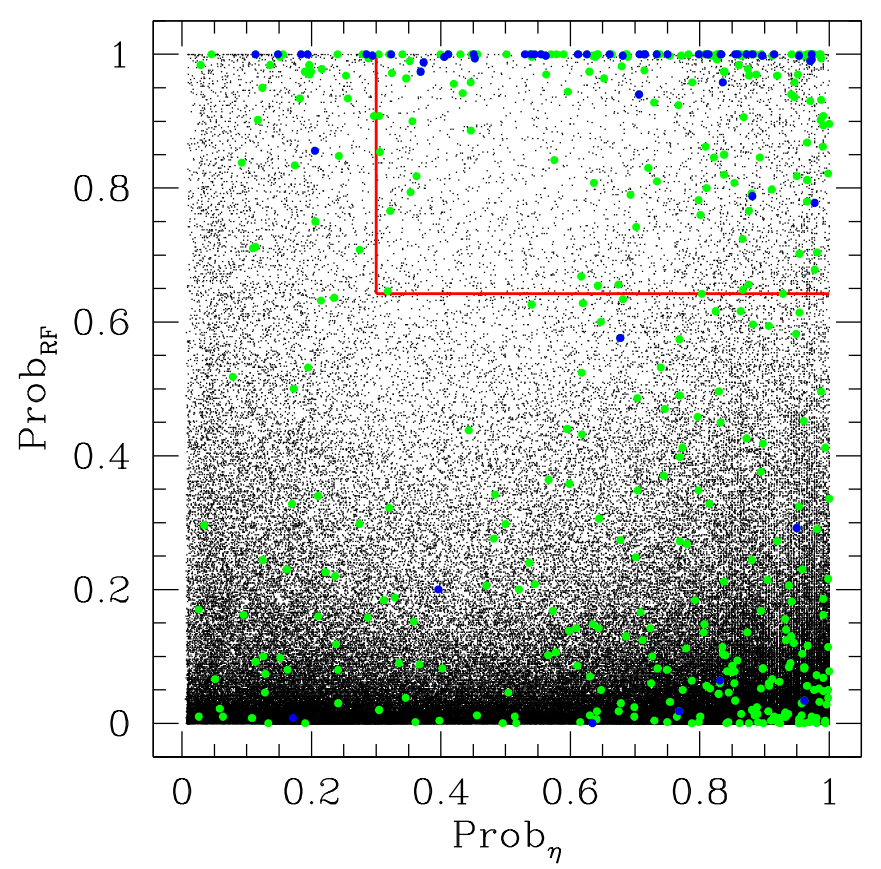}
\includegraphics[width=0.49\columnwidth]{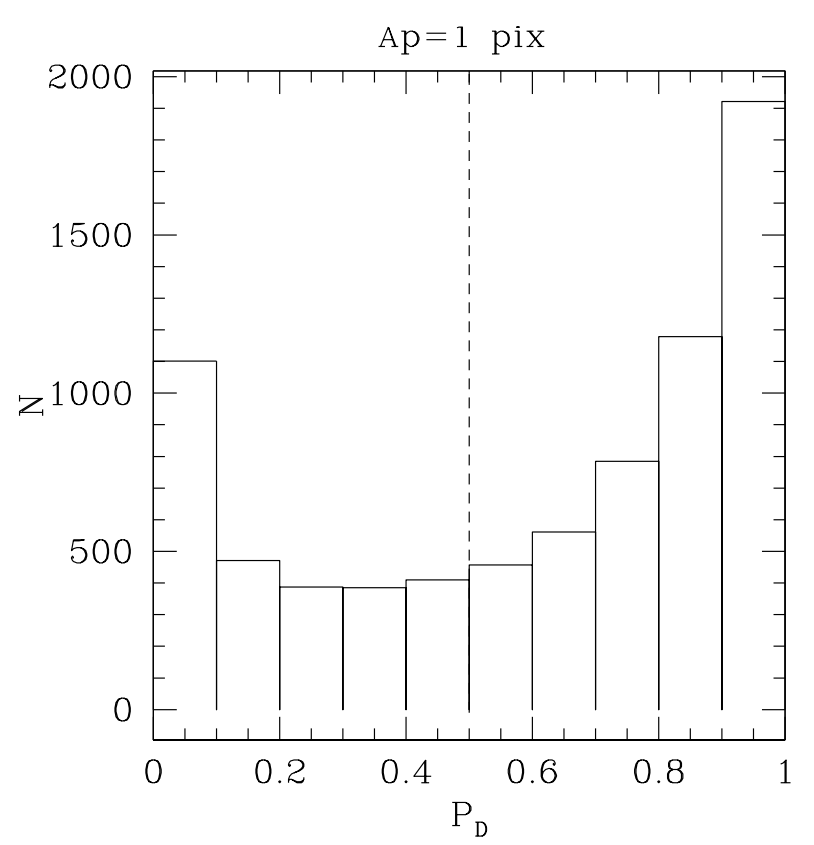}
\includegraphics[width=0.48\columnwidth]{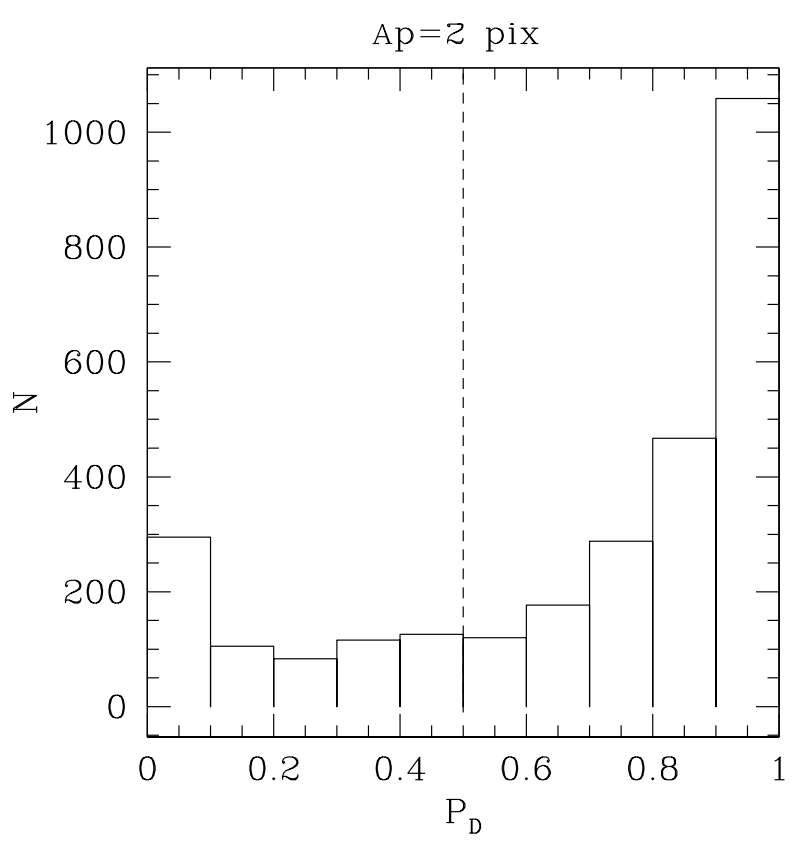}
\caption{
Selection of planetary candidates.
The top figures show the P$\rm_{RF}$ vs P$\rm_{\eta}$ diagrams for lightcurves extracted with
aperture=1 pix (left) and aperture=2 pix (right). Targets selected for the subsequent analysis are located within the top right
rectangles delimited by the red lines.
The bottom histograms show the distribution of P$\rm_D$ probabilities for targets passing the selection in P$\rm_{RF}$ vs P$\rm_{\eta}$.
Green dots denote known TOIs, while blue dots known transiting planets from the TEPcat compilation of \citet{southworth2011}.
}
\label{fig:Prob_RF_ETA}
\end{figure}

\section{Selection of planetary candidates}
\label{sec:selection_planetary_candidates}

To select planetary candidates we started calculating their RFC probability (P$\rm_{RFC}$). We then retrieved, from the {\it Gaia} archive,  the list of contaminant stars of each target out of a distance of 3 arcmin from each target and calculated the probability P$\rm_{\eta}$ described in Sect.~\ref{sec:centroid_motion}, related to the stellar field density. Fig.~\ref{fig:Prob_RF_ETA} (top panels) shows the P$\rm_{RFC}$ vs P$\rm_{\eta}$ diagrams for all the target stars of aperture 1 (left) and aperture 2 (right). For aperture 1 the entire target list (976 814 stars) was used and for aperture 2 we restricted the analysis to the brightest stars (with {\it TESS} magnitude {\it T}$<$11, 201 510 stars). The continuous red lines in the same figures denote our adopted detection thresholds. The RFC thresholds have been described in Sect.\ref{sec:classification}, while for P$\rm _{\eta}$ we adopted a 30$\%$ probability threshold. The number of objects that remained after applying these selection criteria was equal to 7658 stars for aperture 1 and 2836 stars for aperture 2.
Green dots in Fig.~\ref{fig:Prob_RF_ETA} denote known TOIs, while blue dots known transiting planets from the TEPcat compilation of \cite{southworth2011} and will be discussed in Sec.~\ref{sec:discussion}.
For each one of these stars we then calculated the probability P$\rm_{D}$ related to the centroid motion, as described in Sec.\ref{sec:centroid_motion}.  Fig.~\ref{fig:Prob_RF_ETA} (bottom panels)
shows the distribution of candidates
that passed the P$\rm_{RFC}$, P$\rm_{\eta}$ thresholds as a function of the probability P$\rm_{D}$. We applied a 50$\%$ detection threshold for P$\rm_{D}$ as indicated by the vertical dashed lines in the same figure.
The number of candidates that passed this criterium was equal to 4904 stars for aperture 1 and to 2111 stars for aperture 2. 

The whole set of criteria applied up to this point served to identify the candidates with the highest probability of being a transiting planet accordingly to the morphological analysis performed by the RFC,  located in the most isolated stellar fields, and having the smallest centroid motion. We then proceeded with a more detailed scrutiny of each individual candidate. For convenience, we merged the lists of candidates of the two apertures into a single unique list. We then analyzed separately the photometry of each object first in aperture 1 and then in aperture 2. As a first step, we fit a transit model to each detected event, as described in the next Section. 
 
\subsection{Transit analysis}
\label{sec:transit_analysis}

All lightcurves that passed the selection criteria previously described were analyzed using the \cite{mandel2002} transit model algorithm. We assumed a circular orbit, a rectified path across the transit window and that the mass of the transiting body was much smaller than the mass of the primary. We fit for the instant of transit minimum, the transit duration, the planet to star radius ratio, the stellar density, the linear and quadratic limb darkening coefficients and a constant multiplicative factor of the form (1+const). The solutions were obtained using a Levemberg-Marquardt algorithm (LM).
The initial guess parameters for the LM algorithm were obtained from the BLS results and the stellar parameters from the TIC catalog. Limb darkening coefficients  were guessed using the \cite{espinoza2015} software, plugging in the {\it TESS} response function. 
In some cases, we decided to fix some of the fitting parameters (usually the limb darkening coefficients).
After obtaining the results of the analysis, we split the transits into odd and even and performed again the fitting analysis on these two groups of transits separately. In this case we fixed all parameters to their best fit values, with the exception of the radius ratio and the time of minimum. 
Thanks to that a more refined estimate the odd/even transit depth variability was calculated. 

\begin{figure}
\includegraphics[width=\columnwidth]{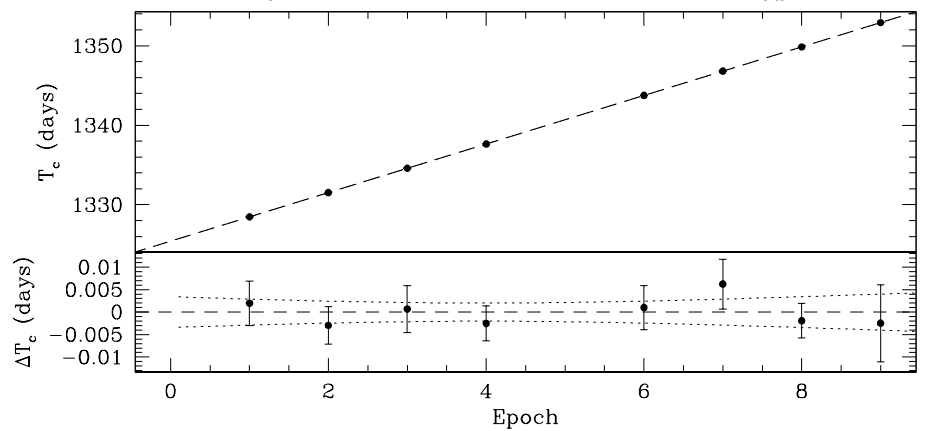}
\caption{
The top diagram shows the measured central transit times (dots) vs the epoch for the candidate 6612853122929259264 ({\it Gaia} DR2 ID). The bottom diagram shows the residuals after subtraction of the best fit linear ephemerides (denoted by the dashed line in the upper diagram).
The upper and lower confidence intervals of the central transit times defined in Eq.~\ref{eq:ephem} are represented by the dotted lines in the bottom diagram.
}
\label{fig:ephem_analysis}
\end{figure}

\noindent
We then proceeded by fitting individually each single transit event. In this case, we fixed all parameters to the values of the best fit model but the instant of transit minimum. The times of the individual transits (T) we obtained were then interpolated with a linear model in order to obtain the best fit linear ephemerides

\begin{equation}
{\rm T}=\widehat{T}_0+\widehat{P}\,\times\,{\rm E}
\end{equation}\noindent
where $\widehat{T}_0$ is the time of the reference epoch and $\widehat{P}$ is the orbital period and E is the epoch. 
The null epoch is set by definition as the epoch before the first observed transit.
By assuming normally distributed errors, we also calculated the uncertainties on $\widehat{P}$ and $\widehat{T}_0$ 

\begin{equation}
\Delta \widehat{P}=\sqrt{\frac{\frac{1}{{\rm N}-2}\sum_{i=1}^{{\rm N}} \widehat{\varepsilon}^2_i }{\sum_{i=1}^{{\rm N}} (\rm E_i-\bar{E})}}
\end{equation}
\begin{equation}
\Delta \widehat{T}_0=\Delta \widehat{P} \sqrt{\frac{1}{{\rm N}}\sum_{i=1}^{{\rm N}} \rm E^{2}_{i}}
\end{equation}\noindent
where $\widehat{\varepsilon}_i$ are the residuals of the fit, $\rm E_i$ and $\rm \bar{E}$  are the i-th epoch and the average epoch and $\rm N$ is the number of transits. We also estimated the confidence bands around the regression line as

\begin{equation}
(T_0 + P \,{\rm E}) \in \left[ \,\widehat{T_0} + \widehat{P} \,\rm E \pm \{\alpha+\beta\times E\}\right]
\label{eq:ephem}
\end{equation}

\noindent
where 

\begin{equation}
{\rm \alpha}= t^*_{{\rm N} - 2} \sqrt{ \left(\frac{1}{{\rm N} - 2} \sum\widehat{\varepsilon}_i^{\,2} \right) \cdot \left(\frac{1}{{\rm N}}\right)}-
 t^*_{{\rm N} - 2} \bar{E} \sqrt{ \left(\frac{1}{{\rm N} - 2} \right) \cdot \left(\frac{\sum\widehat{\varepsilon}_i^{\,2}}{\sum ({\rm E_{i}} - \bar{\rm E})^2}\right)}
 \label{eq:alpha}
\end{equation}

\begin{equation}
{\rm \beta}= t^*_{{\rm N} - 2} \sqrt{ \left(\frac{1}{{\rm N} - 2} \right) \cdot \left(\frac{\sum\widehat{\varepsilon}_i^{\,2}}{\sum ({\rm E_{i}} - \bar{\rm E})^2}\right)}
\label{eq:beta}
\end{equation}

\noindent
and $t^*_{\rm N - 2}$ is the (1-$\frac{\gamma}{2}$)-th quantile of the Student's t-distribution with N-2 degrees of freedom. We adopted $\gamma$=0.05 that corresponds to a 95$\%$ confidence level. Eq.\ref{eq:ephem} permits a straightforward estimate of the uncertainty of transit times at any arbitrary epoch (E$\ge$0).
An example is shown in Fig.~\ref{fig:ephem_analysis}, where the 95\% confidence interval is represented by the dotted lines in the bottom panel.

\begin{figure*}
\includegraphics[width=17cm]{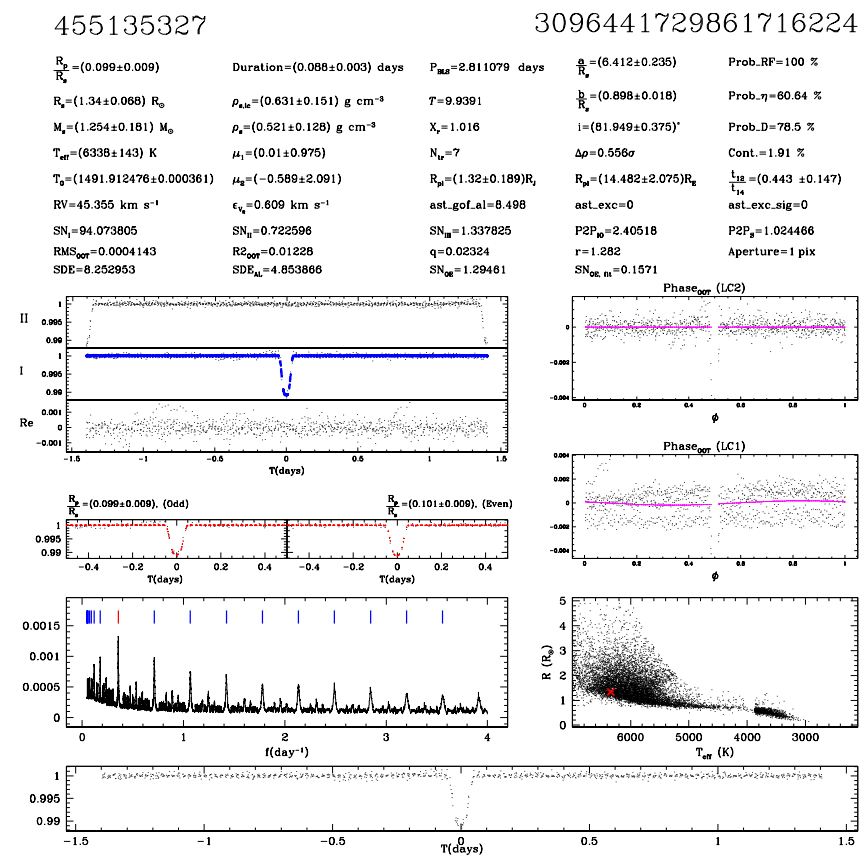}
\caption{
Example of diagnostic plot used during the screening process of the candidates. The object represented is HAT-P-30/WASP-50.
}
\label{fig:diagnostic}
\end{figure*}

\subsection{Individual Object Analysis}
\label{sec:individual_object_analysis}

After the results of the fit analysis were obtained, we scrutinized each object individually. For each candidate, we produced two diagnostic figures where we summarized all the information known up to this point of the analysis. The first plot is illustrated in Fig~\ref{fig:diagnostic}
for the case of HAT-P-30/WASP-50. It presents several useful diagrams including the phase folded lightcurve with the best fit model and the residuals of the fit, the folded lightcurve at secondary phase, out-of-transit diagrams to determine the presence of out-of-transit variability, a stellar radius vs effective temperature diagram, the BLS periodogram, the odd/even diagrams with the corresponding fits in addition to several quantitative measurements related to the candidate, the star, the classification algorithm, and the dynamical constraints.
The second plot is entirely similar to the ones presented in Fig~\ref{fig:targ_cont} and was used to analyze the centroid motion distributions. The meaning of all diagrams and quantities reported in these figures is explained in the documentation released with the data products.
Both aperture photometries were inspected together with the centroid motion diagram. We searched in particular for differential transit depths in the two apertures. In some cases we accepted some ambiguous situations (e.g. some transits detected only in one aperture, or Mahalanobis rank failures when the probability thresholds are met, etc.) and highlighted them with a bitmask flag in our catalog (see Table~\ref{tab:catalog}). For all objects which were not obviously spurious we then further refined the transit fit analysis reported in Sect.$\ref{sec:transit_analysis}$. This time an
LM algorithm with bootstrap analysis with 1000 iterations was performed to obtain the best fit-parameters and their uncertainties.

\begin{figure*}
\includegraphics[width=\columnwidth]{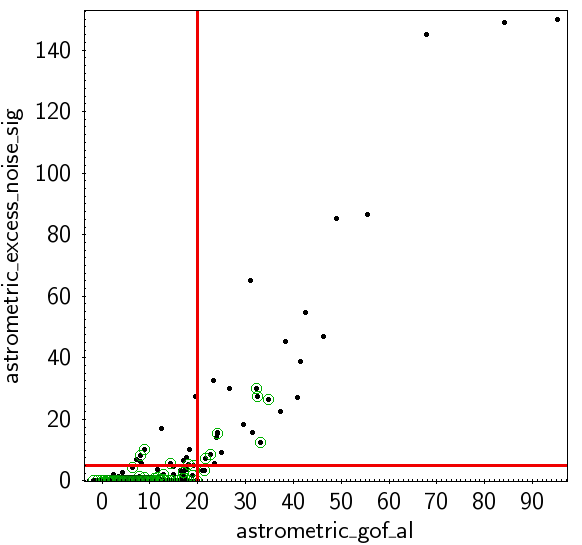}
\includegraphics[width=\columnwidth]{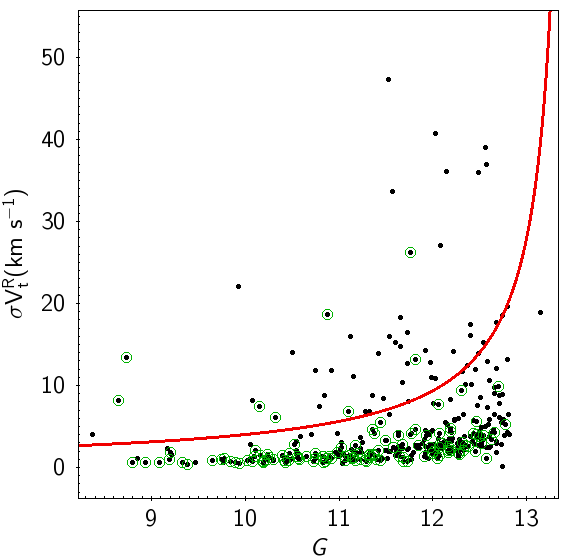}
\caption{
{\it Left:} 
Significance of the Astrometric excess Noise vs Astrometric Goodness of Fit in the Along-Scan direction from the {\it Gaia} DR2 archive relative to the planetary candidates discussed in this work.
{\it Right:} Planetary candidates' standard deviation of {\it Gaia} radial velocity measurements against {\it Gaia} $G$-band magnitude. The red lines denote selection criteria we used to flag potential binaries (see text), while green circles show known TOIs and CTOIs.
}
\label{fig:dynamical_constraints}
\end{figure*}

\subsection{Dynamical constraints}

By using {\it Gaia} DR2 we determined the possible presence of binaries in our sample, as described in Sec.~\ref{sec:dynamical_constraints}. In Fig.~\ref{fig:dynamical_constraints} we show the candidates' significance of the astrometric excess noise vs the astrometric goodness of fit diagram (left) and the standard deviation of the radial velocity measurements vs the {\it G}-band diagram (right). Among the candidates 51 show exceedingly large radial velocity standard deviations, 27 significant astrometric excess and 8 of them both.
The total list of suspected binaries, based on our adopted thresholds, is therefore equal to 70 objects. In Fig.~\ref{fig:dynamical_constraints} (left), we represented with green open circles stars which are either known TOIs ({\it TESS} objects of Interest) or known CTOIs (Community TOIs). Eight of these stars are in the list of suspected binaries, based on their RV standard deviations. They are: star 260130483 (TOI 933.01), 207081058 (TOI 948.01), 9033144 (TOI 367.01), 49899799 (TOI 416.01), 219345200 (TOI 706.01), 382068562 (TOI 924.01), 423670610 (TOI 850.01) and 146438872 (TOI 948.01).
Seven TOIs/CTOIs are instead flagged on the basis of the astrometric indicators (Fig.~\ref{fig:dynamical_constraints}, right): star 122612091 (TOI264.01, which is a known planet, WASP-72)
257567854 (TOI403.01/WASP-22),
40083958 (TOI851.01),
183593642 (TOI355.01),
261261490, 429302040 (TOI1905.01/WASP-107b) and 
219345200 (TOI706.01). This last object is common to both lists.

\begin{figure*}
\includegraphics[width=18cm]{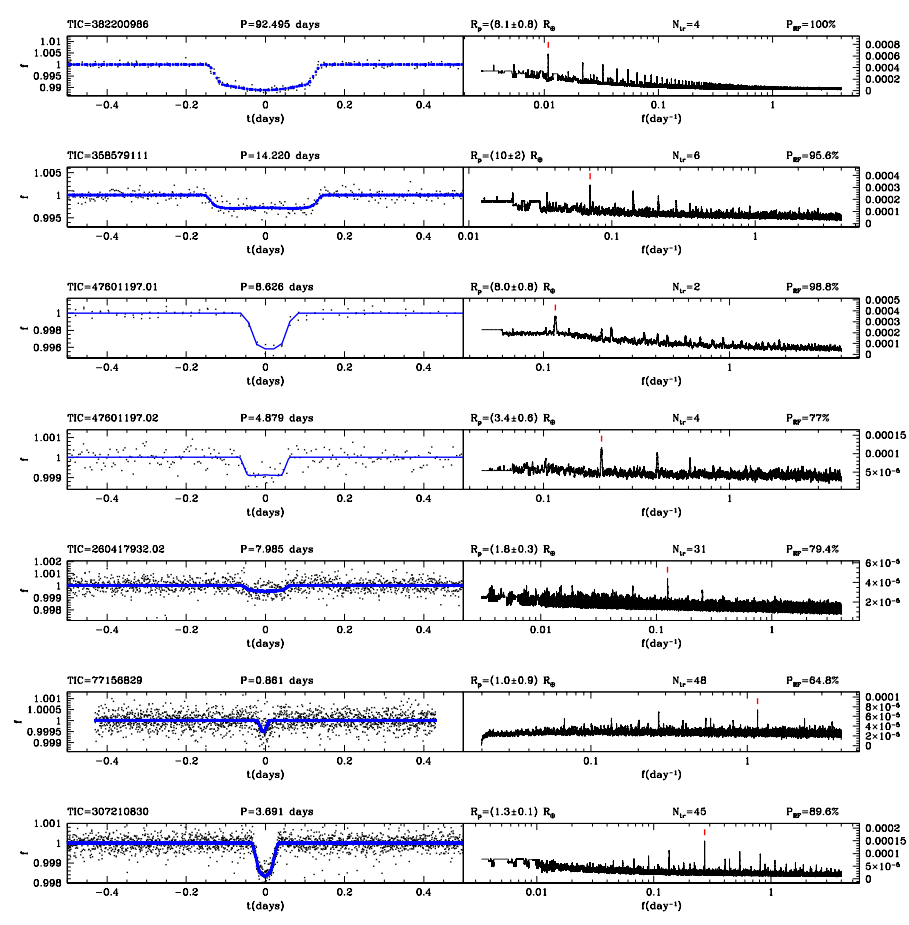}
\caption{
A few representative examples of transiting planetary candidates we detected. On the left the folded lightcurves and on the right the corresponding BLS periodograms. The labels on the top report the TIC ID number, the orbital period (P), the transiting object's radius (R$\rm_{p}$), the number of observed transits (N$\rm_{tr}$) and the random forest probability (P$\rm_{RF}$).
}
\label{fig:lcfig}
\end{figure*}

\subsection{Cross-match with external catalogs}
\label{sec:crossmatch}

We checked if the list of objects we found was included in known lists of variables or false positives. We cross-matched our catalog with the International Variable Star Index catalog searching for all variables within a distance of 3 arcmin from our targets and we found 60 stars. Almost all of them have also a reported period. By imposing a  precision of 1\% between the period we found and the VSX period (or half or twice this period), the list is restricted to 30 objects. All of them are known planets from the WASP, HAT and HATS surveys, with the exception of one object which corresponded to star 1SWASPJ055532.69-571726.0 in VSX and to star 734505581 in our catalog. Such object is reported to be a detached eclipsing binary and it was therefore eliminated from the list. The matched objects without a reported period were SN 1995V, a SN II Supernova sitting 159 arcsec from 35857242, the variable ASASSN-V J071237.50-530912.6 at about 54 arcsec from 344087362, NSV 4303 at 0.2 arcsec from 13737885 which is classified as CST (that is a retracted variable star) and CR Gru at 171 arcsec from 265612438 classified as LB, a slow irregular variable of late spectral type. 

\noindent
We then considered the KELT Follow-Up Network and Transit False Positive Catalog \citep[KELT-FUN,][]{collins2018}, an all-sky catalog of 1128 bright stars (6 < {\it V} < 13) showing transit-like features in the KELT light curves which have been then ruled out as false positives by follow-up observations. By repeating the same procedure reported above we found one match with our list, which corresponds to star 144426921 (TIC ID). This object is classified as an SB2 (multi-lined binary) and it is has been therefore eliminated from our list.

\noindent
Another valuable compilation of known false positives is the SuperWASP dispositions and false positive catalogue \citep{schanche2019b}
which lists 1041 Northern hemisphere SuperWASP targets, rejected as false positives by follow-up observations. Part of our targets are present also in the equatorial region and northern sky. We found in this case 3 matches with our catalog, which correspond to stars 443618156, 16490297 and 9727392 (TIC IDs). These objects are classified as EB or EBLM binaries and were therefore eliminated.
We note that star 9727392 is also included in the list of known TOIs (TOI236.01). The public comment reports a 1700 ppm secondary detection and flagged it as a likely EB.

\section{Results}
\label{sec:results}

The search yielded 396 candidates among which 144 are known TOIs or community TOIs and 252 are new candidates\footnote{We refer to the candidates' relase of June 19, 2020.}. 
We also compared our candidates' list with the SPOC multisector TCE list available from MAST
\footnote{
\url{http://archive.stsci.edu/tess/bulk_downloads/bulk_downloads_tce.html} the relevant list is tess2018206190142-s0001-s0013\_dvr-tcestats.csv} and found 22 matches among which 20 were known TOIs or CTOIs. The other two stars are star 141770592 (TIC ID) and 177350401.
Table~\ref{tab:quartiles_porb_rad} reports the minimum, the first quartile, the median, the mean, the third quartile and the maximum values of the period and radius distributions. The median values of the distributions correspond to Jupiter planets in short orbital period (Hot Jupiters). The radii distribution is extended down to 1$\rm R_{\oplus}$ and the orbital period distribution up to $\sim$105 days. 
By considering the distribution of the candidates' impact parameters (\textit{b}) obtained from the fitting analysis we found that $\sim$50$\%$ (197) of the candidates have \textit{b}$\le$0.8. Because of the long temporal cadence of TESS FFIs, planetary candidates, especially around late
type dwarfs, may have preferentially V-shaped transits. Nevertheless, the statistical argument for the planetary nature of b$<$0.8 objects is generally stronger than for b$>$0.8 objects \citep[e.g.][]{seager2003}. To facilitate follow-up analysis and target prioritization, we subdivided our list of candidates in five tiers as a function of the impact parameter value, with candidates having \textit{b}$\le$0.2 belonging to the first tier (see also Table~\ref{tab:catalogcsv}).

\noindent
We discovered a long period candidate planet (P$\sim$92 days) with a radius of about 0.7 R$\rm_J$ around a likely early K dwarf (TIC ID =382200986).  
We recovered a Jupiter like candidate around star 309792357. We note that this object is included in the list of TOIs (TOI ID=199.01). The public comments indicate the presence of a single transit, but we recovered three transits and inferred a period of $\sim$105 days. According to our analysis therefore, this object is a long period transiting candidate.
We also confirmed the presence of a long-period candidate (P$\sim$89 days) around a late G sub-giant star (TIC=260130483, TOI ID=933.01) and a long period (P$\sim$53 days) giant planet candidate around star with TIC=350618622 (TOI ID=201.01). Star 270341214 (TOI173.01) is another interesting object. The public comment of this TOI indicates the presence of a single transit. We detected however 2 transits, one in sector 1 and one in sector 13. The BLS period is $\sim25$ days likely because of a saturation effect on the periodogram due to the presence of a large gap in the data. Indeed, this star has been observed only in these two sectors and therefore this object could have potentially a very long period ($\sim$ 327 days). Because of the large gap in the data, it is entirely possible that the period is a sub-multiple of this value. We note that the transit duration is  also long (8.42 hr) and the star radius is $\sim$1.4 R$_{\odot}$. This somewhat supports the fact that this object could be a long period planetary candidate. Moreover, we additionally detected 15 single transit events (one of which is known, TOI 706.01, TIC=219345200) which may potentially be long period candidates.

\noindent
We detected a conspicuous population of candidates with orbital periods between 10 days and 50 days and radii between 2 R$_{\oplus}$ and 2.6 R$\rm_{J}$. In total this sample amounts to 64 objects, 42 of which are new candidates. There are 39 candidates with radius R$\rm_{p}<$ 4 $\rm R_{\oplus}$, where 15 are new candidates.

\subsection{Multiples}
We searched for multiple planetary candidates after subtracting the best fit model of the primary candidate and repeating the transit search as described in the previous sections.
We adopted the same detection tresholds adopted for the primary transits. In this way we detected a new candidate around star 47601197 with radius R$\rm_{p}$=(3.4$\pm$0.6) R$\rm_{\oplus}$
and period equal to $\sim4.9$ days, nearly half the one of the primary candidate (which is itself a new candidate with P=8.6 days, R$\rm_{p}$=9.0$\pm$1.0 R$\rm_{\oplus}$).
We confirmed the super-Earth around star 259377017  \citep[TOI ID=270.02,][]{gunther2019} with
period P$\sim$11.4 days and radius R$\rm_{p}$=(2.2$\pm$0.3) R$\rm_{\oplus}$\footnote{We detected also the signal of TOI 270.03 on aperture=2 pix with period $\sim$3.4 days and radius $\sim$1.5 R$_{\oplus}$, but the random forest probability was 52.8$\%$, below our detection threshold.}.
We found a new candidate around star
260417932 with period P$\sim$8 days and R$\rm_{p}$=(1.8$\pm$0.3) R$\rm_{\oplus}$.

Fig.~\ref{fig:lcfig} shows some representative examples of candidates we detected. Along with the folded lightcurves (on the left) and the BLS periodograms (on the right) we report the TIC ID number, the orbital period (P), the radius of the transiting object (R$\rm_{p}$), the number of observed transits (N$\rm_{tr}$) and the random forest probability (P$\rm_{RF}$).

\begin{table}
	\centering
	\caption{Minimum, first quartile (1$\rm ^{st}$Q), median, mean, third quartile (3$\rm ^{rd}$Q) and maximum values of candidates' orbital periods (P) and radii (R$\rm_{p}$) distributions.}
	\label{tab:quartiles_porb_rad}
	\begin{tabular}{ccccccc} 
    \hline
	  & Min & 1$\rm ^{st}$Q & Median & Mean & 3$\rm ^{rd}$Q & Max \\	
	\hline	
	R$\rm_{p} (R_{\oplus})$ & 1.0 & 7.0 & 11.5 & 11.8 & 15.1 & 28.9 \\
	
	P (days)& 0.25 & 2.1 & 3.8 & 6.6 & 7.2 & 104.9 \\
		\hline
	\end{tabular}
\end{table}

\begin{figure}
\includegraphics[width=\columnwidth]{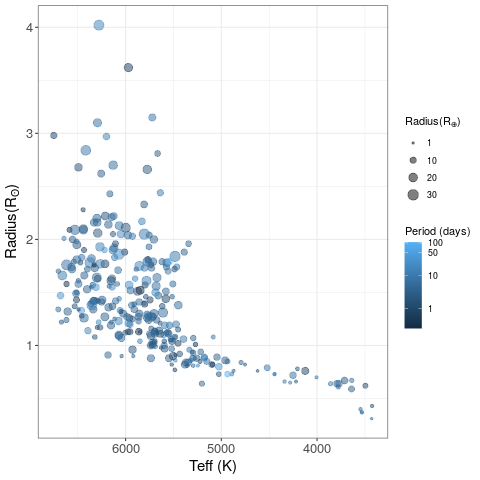}
\caption{
Stellar radius vs T$\rm_{eff}$ for planetary candidates' host stars. The different sizes and colors of the points in the diagram codify the planetary candidates' radii and orbital periods as explained in the legend.
}
\label{fig:teff_radius}
\end{figure}

\begin{figure}
\includegraphics[width=\columnwidth]{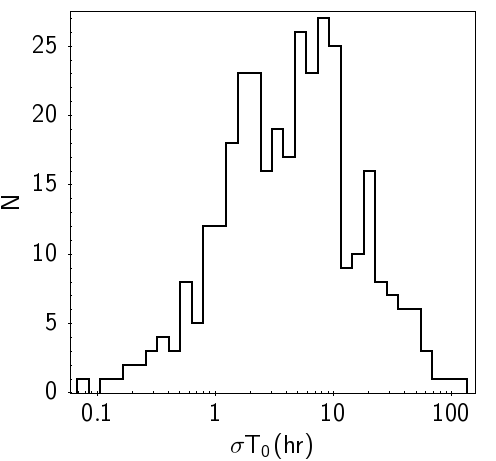}
\caption{
Uncertainty on the central transit time of planetary candidates calculated with Eq.~\ref{eq:ephem}, assuming all objects will be observed on August 1, 2020.
}
\label{fig:ephem}
\end{figure}

\section{Discussion}
\label{sec:discussion}

Fig.~\ref{fig:teff_radius} shows the stellar radius vs effective temperature diagram of the transiting candidates' host stars. The dimensions of the points and their colors denote the planetary radii and periods, as reported in the legend. We plot only stars for which the relative errors on the planetary radius is smaller than 30$\%$
and excluded single transit hosts\footnote{An error of 10$\%$ was added in quadrature to the stellar radius error (and propagated to the planetary radius error) for stars with T$\rm_{eff}<3950$ K because the error reported in the TIC catalog was very small.}. In total the diagram represents 304 candidates. From this figure we can deduce that candidates are present in all the portions of the diagram, from cool late type dwarfs to evolved sub-giants stars.

It's important to keep in mind some caveats once interpreting the results discussed in this work. First, our analysis has been purposely limited to less crowded regions of the sky by imposing a cut-off threshold on the local average stellar density. This is a reasonable choice for a work like this one, since it increases the chances of correct source identification. Unavoidably however, as demonstrated in Fig.~\ref{fig:Prob_RF_ETA}, this likely reduced the efficiency of the search. We found 618 TOIs in our initial target list in aperture 1 (436 in aperture 2) and 139 were eliminated on the basis of the stellar density condition, which gives a fractional loss of 20$\%$, 30$\%$ for aperture 1 and 2, respectively. If we consider the TEPcat catalog \citep{southworth2011} we found 118 stars in common with aperture 1 sample (56 for aperture 2). Out of them 28 and 16 were eliminated by the condition on the stellar density which gives a fractional loss of 23$\%$ and 29$\%$ for aperture 1 and 2, respectively. Then, by restricting the analysis to only those targets that passed the stellar density condition, we obtain that the overall TOIs recovery rate is 29$\%$-42$\%$ and that the TEPcat recovery rate is 57$\%$-83$\%$. The gap between these two samples' recovery rates can be explained by the fact that the TOI sample is strongly biased towards small radii objects. In the TOI sample in common with our target list 48$\%$ of the candidates have a radius $<5\,$R$_{\oplus}$, whereas in the TEPcat sample this percentage decreases to 10$\%$. As shown in Fig.\ref{fig:efficiency}, we expected a significant drop in our efficiency in this regime. 
It should be also recalled that our detection thresholds were set to relatively conservative values, as demonstrated for example by the ROC curves of Fig.~\ref{fig:ROC}. There would be certainly room to increase the detection efficiencies in both apertures by tweaking the detection thresholds, at the price however of tolerating higher false positive rates than the 1$\%$ level we decided to adopt. Certainly other factors impact efficiencies beyond the ones we considered. A complete examination of them goes well beyond the purpose of this work, but it is sufficient to recall for example that our training algorithm was based on simulated datasets and that these datasets certainly are not exhaustively incorporating all the possible sources of false positives which may affect the detection of transiting planets. Despite these limitations, it is remarkable that we managed to discover and to recover a substantial sample of small size and of long period candidates, as demonstrated in Sec.~\ref{sec:results}. Because our detection efficiencies are clearly reduced in these two domains (e.g. Fig~\ref{fig:efficiency}) this clearly points towards the conclusion that the abundance of these objects is large. 

As a last point, it is also worth to have a look at the ephemerides' uncertainties of the detected objects. We calculated the uncertainty in the central transit time as the difference between the maximum and minimum ephemerides obtained from by Eq.~\ref{eq:ephem}, assuming to observe on August 1, 2020. Fig.~\ref{fig:ephem} shows the distribution of the uncertainties which range from a minimum of 5 min to a maximum of 4.8 days with a median value of 4.7 hours. This fact is in itself a very good reason for an extended mission to observe again these objects and refine their ephemerides.
For example, \cite{bouma2017} showed that if we detect an additional transit 1 year after the final observed transit from the Primary Mission, the uncertainty on the mid-transit time decreases by an order of magnitude. This will permit to observe several candidates for many years to come.

\section{Conclusions}
\label{sec:conclusions}

In this work, we explored the potential of {\it TESS} FFIs to detect transiting planets around a well defined sample of FGKM dwarfs and sub-giant stars in the Southern ecliptic hemisphere. We discussed our new reduction pipeline, DIAmante, based on the differential imaging analysis
to optimally extract multi-sector photometry from FFIs. We then presented our post-correction analysis method and transit search approach. In particular, we discussed a morphological classification algorithm based on the Random Forest technique which permits to discriminate planetary transiting candidates from different categories of false positives. We discussed the ROC analysis and the overall algorithm performances. We then focused our attention on the centroid motion algorithm, introducing different quantitative metrics to reduce the chances of wrong source identification. First, we discussed the Mahalanobis distance classification, which is based on the distance of the target star from the centroid motion distributions derived in four concentric apertures centered on the target. We then developed a probabilistic model to account for the probability to misidentify a source of variability when the local average stellar density increases. We showed that when the local stellar average number density is equal to 0.84 stars pix$^{-2}$, the probability of correct source identification is 50$\%$. We then analyzed the empirical behaviour of different quantities related to the centroid motion distributions, combining them into a unique metric. We showed that when the combined effect of the centroid motion disposition and its uncertainty amount to about 7.2 arcsec (that is 34$\%$ of the {\it TESS} average pixel dimension), the probability that the target source is the source of variability is equal to 50$\%$. We then considered {\it Gaia} dynamical constraints and presented the method of selection of planetary candidates. Our search produced a list of 396 planetary candidates, out of which 252 are new candidates. By applying dynamical constraints from {\it Gaia} we found that 70 (18$\%$) candidates show evidence of unresolved binarity. This list includes 14 TOIs/CTOIs. The planetary radii distribution ranges from 1$\rm R_{\oplus}$ to 2.6 R$\rm_{J}$ with a median value 1 R$\rm_{J}$. The orbital period distribution ranges between $\sim$0.25 days and $\sim$105 days with a median value of 3.8 days. We discovered a new long period candidate with a period of 92 days and R=0.7 R$\rm_{J}$ and inferred a long period for TOI 199.01 (P=$\sim$105 days). Other two TOIs with periods larger than 50 days were detected. We also discussed the curious case of TOI 173.01, observed only in Sector 1 and Sector 13 and for which two transits separated by $\sim$327 days were detected. Additionally, 15 single transit events have been found and could potentially be long period candidates. We found 64 candidates with orbital periods between 10 days and 50 days, 42 of which are newly discovered. We also detected 39 candidates with radii smaller than 4 R$\rm_{\oplus}$, and 15 of them are new. Finally, we found a new multiple system around star 4701197 and found a new candidate planet around star 260417932. 

During our work we largely benefited from the public compilations of variable stars and known false positives which have been made publically available. This is worth mentioning, given that these lists are extremely valuable resources which permit to avoid duplication of efforts and we certainly encourage a broad diffusion of similar lists available to other groups.

All of the objects discussed in this work have been detected by {\it TESS} about one or two years ago. We showed that the median uncertainty of the candidates' central transit time is $\sim$4.7 hours. This is a very good reason for {\it TESS} coming back during its extended mission refreshing their ephemerides.

As time passes, the value of {\it TESS} data increases. Adding novel observations to objects measured in the past reduces the impact of systematics, increases the chances to detect smaller and longer period planets. The importance of {\it TESS} goes well beyond its primary mission goals. The database of high precision lightcurves that is created from
{\it TESS} FFIs and from {\it TESS} imagettes is an important legacy also for the preparation of future planet hunting missions like {\it PLATO}. Nevertheless a wealth of other different studies in nearly any field of astrophysics will benefit of it. What it is sure is that as {\it TESS} will continue now into its extended mission, {\it TESS} data miners will continue to follow its swings around the sky to unveil the hidden gems it will observe. 

With this work, we release the catalog of the 396 candidates we found together with their lightcurves and diagnostic plots. 
This material is submitted to the Mikulsky Archive for Space Telescopes (MAST) under the section High Level Science Products (HLSP) and the project's name DIAmante. The catalog description is reported in Table~\ref{tab:catalogcsv}. 
Newly discovered candidates are also reported in the ExoFOP portal as Community TOIs (CTOIs).

\section*{Acknowledgements}
This paper includes data collected by the { \it TESS} mission. Funding for the {\it TESS} mission is provided by the NASA Explorer Program. This research has made use 
of the International Variable Star Index (VSX) database, operated at AAVSO, Cambridge, Massachusetts, USA. The authors are grateful to the anonymous referee for all the comments and suggestions that permitted to improve the manuscript and increased the visibility of our work.

\section*{Data availability}
The data underlying this article are available in MAST, at
\doi{10.17909/t9-p7k6-4b32} and 
\url{https://archive.stsci.edu/hlsp/diamante} 
and at the ExoFOP portal at \url{https://exofop.ipac.caltech.edu/tess/}.




\bibliographystyle{mnras}
\bibliography{MMontalto} 





\begin{table*}
	\centering
	\caption{Content of the catalog released with this work.}
	\label{tab:catalogcsv}
	\begin{tabular}{clcl} 
    \hline
	Column number & Column name  & units & Description \\	
	\hline
	1 & ticID & - & TIC catalog ID number \\
	2 & gaiaID & - & {\it GAIA} DR2 source ID number \\
	3 & ra & deg & Right ascension from {\it GAIA} DR2\\
	4 & dec & deg & Declination from {\it GAIA} DR2\\
	5 & p & - & Radius ratio \\
	6 & pErr & - & Error on radius ratio \\
	7 & t0Fit & days & Central transit time from transit fit \\
	8 & t0FitErr & days & Error on central transit time from transit fit\\
	9 & periodBLS & days & Orbital period from BLS analysis \\
	10 & duration & days & Transit duration \\
	11 & durationErr & days & Error on transit duration \\
	12 & denFit & g cm$^{-3}$ & Stellar density from transit fit \\
	13 & denFitErr & g cm$^{-3}$ & Error on stellar density from transit fit \\
	14 & u1 & - & Linear limb darkening coefficient \\
	15 & u1Err & - & Error on linear limb darkening coefficient \\
	16 & u2 & - &  Quadratic limb darkening coefficient\\
	17 & u2Err & - & Error on quadratic limb darkening coefficient \\
	18 & const & - & Constant term in transit fit \\
	19 & constErr & - & Error on constant term in transit fit \\
	20 & chir & - & Reduced chi squared of the transit fit \\
	21 & deltarho & - & Difference of stellar density from the TIC and from transit fit, normalized by the square sum of the errors\\
	22 & trdepth & ppm & transit depth \\
	23 & rmsoot & - & Root mean square of out of transit measurements \\
	24 & ar & - & Ratio of semi-major axis to stellar radius \\
	25 & arErr & - & Error on ratio of semi-major axis to stellar radius \\
	26 & b & - & Impact parameter \\
	27 & bErr & - & Error on impact parameter \\
	28 & i & deg & Inclination \\
	29 & iErr & deg & Error on inclination \\
	30 & t12t14 & - & Ratio of ingress time to total transit time\\
	31 & t12t14Err & - & Error on ratio of ingress time to total transit time\\
	32 & rpj & R$\rm_{J}$ & Candidate radius \\
	33 & rpjErr & R$\rm_{J}$ & Error on candidate radius  \\
	34 & rpe & R$\rm_{\oplus}$ & Candidate radius \\
	35 & rpeErr & R$\rm_{\oplus}$ & Error on candidate radius \\
	36 & rank1 & - & First Mahalanobis rank \\
	37 & rank2 & - & Second Mahalanobis rank \\
	38 & rank3 & - & Third Mahalanobis rank \\
	39 & rank4 & - & Fourth Mahalanobis rank \\
	40 & prf & \% & Random Forest probability \\
	41 & pd & \% & Centroid motion probability (Eq.~\ref{eq:pd}) \\
	42 & peta & \% & Local average stellar number density probability (Eq.~\ref{eq:peta}) \\
	43 & t0Lin & days & Central transit time from linear fit\\
	44 & t0LinErr & days & Error on central transit time from linear fit\\
	45 & periodLin & days & Orbital period from linear fit\\
	46 & periodLinErr & days & Error on orbital period from linear fit\\
	47 & alpha & days & $\rm \alpha$ parameter from Eq.~\ref{eq:alpha}\\
	48 & beta & days & $\rm \beta$ parameter from Eq.~\ref{eq:beta}\\
	49 & teff & K & Stellar effective temperature from TIC\\
	50 & teffErr & K & Error on stellar effective temperature from TIC\\
	51 & radius & R$\rm_{\odot}$ & Stellar radius from TIC\\
	52 & radiusErr & R$\rm_{\odot}$ & Error on stellar effective temperature from TIC\\
	53 & mass & M$\rm_{\odot}$ & Stellar mass from TIC\\
	54 & massErr & M$\rm_{\odot}$ & Error on stellar mass from TIC\\
	55 & rho & g cm$^{-3}$ & Stellar density from TIC\\
	56 & rhoErr & g cm$^{-3}$ & Error on stellar density from TIC\\
	57 & contratio & \% & Contaminatio ratio from TIC \\
	58 & tmag & - & {\it TESS} magnitude from TIC\\
	59 & vmag & - & {\it V} magnitude from TIC\\
	60 & gmag & - & {\it G} magnitude from TIC\\
	61 & rv & km s$^{-1}$ & Radial velocity from {\it Gaia} DR2\\
	62 & rvErr & km s$^{-1}$ & Error on radial velocity ($\rm\epsilon_{RV}$) from {\it Gaia} DR2\\
	63 & nrv & - & Number of eligible transits used to derive the median radial velocity from {\it Gaia} DR2\\
	64 & srv & km s$^{-1}$ & Standard deviation of radial velocity measurements \\
    \hline
    \label{tab:catalog}
	\end{tabular}
\end{table*}

\addtocounter{table}{-1}
\begin{table*}
	\centering
	\caption{ - Continued.}
	\label{tab:catalogcsv}
	\begin{tabular}{clcl} 
    \hline
	Column number & Column name  & units & Description \\	
	\hline
	65 & gofAl & - & Goodness of fit in the along scan direction from {\it Gaia} DR2\\
	66 & astroExcess & mas & Excess of astrometric noise from {\it Gaia} DR2\\	
	67 & astroExcessSig & - & Significance of Excess of astrometric noise from {\it Gaia} DR2\\
	68 & snI & - & S/N ratio of primary transit from Eq.~\ref{eq:SNI}\\
	69 & snII & - & S/N ratio of secondary eclipse from Eq.~\ref{eq:SNII}\\
	70 & snIII & - & S/N ratio of tertiary eclipse from Eq.~\ref{eq:SNIII}\\
	71 & snOE & - & Odd/Even S/N ratio from Eq.~\ref{eq:SNOE}\\
	72 & snOEFit & - & Odd/Even S/N ratio from transit fit \\
	73 & r2oot & - & $\rm R^{2}_{OOT}$ parameter from Eq.~\ref{eq:R2OOT}\\
	74 & q & - & Fractional transit duration from Eq.~\ref{eq:q}\\
	75 & sde & - & Signal Detection Efficiency from Eq.~\ref{eq:SDE}\\
	76 & sdeAL & - & Signal Detection Efficiency of Alias peaks from Eq.~\ref{eq:SDEAL}\\
	77 & p2pio & - & In/Out of transit point-to -point noise from Eq.~\ref{eq:P2PIO}\\
	78 & p2ps & - & P2P$\rm_{s}$ parameter from Eq.~\ref{eq:P2PS}\\
	79 & r & R$\rm_{J}$ & Estimated candidate radius from Eq.~\ref{eq:radius} \\
	80 & ntr & - & Number of transits \\
	81 & apnum & pix & Aperture photometry for which the transit analysis was performed\\
	82 & btier & - & A level flag based on the impact parameter value \\
	   &       &   & btier=1, b$\le$0.2 \\
       &       &   & btier=2, 0.2$<$b$\le$0.4 \\	   
       &       &   & btier=3, 0.4$<$b$\le$0.6 \\       
       &       &   & btier=4, 0.6$<$b$\le$0.8 \\
       &       &   & btier=5, b$>$0.8 \\
	83 & bitmask & - & Bitmask flag\\
	   &         &   & bitmask=1 - binary according to {\it Gaia} DR2 dynamical constraints \\
	   &         &   & bitmask=2 - Single transit \\
	   &         &   & bitmask=4 - At least one Mahalanobis rank failed \\
       &         &   & bitmask=8 - Present only in Aperture 1 \\	   
       &         &   & bitmask=16 - Present only in Aperture 2 \\
    84 & mult & - & Candidate numeration within multiple systems\\
    \hline
	\end{tabular}
\end{table*}

\bsp	
\label{lastpage}
\end{document}